# Towards real-time finite-strain anisotropic thermo-visco-elastodynamic analysis of soft tissues for thermal ablative therapy

Jinao Zhang[1], Remi Jacob Lay[1], Stuart K. Roberts[2], Sunita Chauhan[1]

[1]Department of Mechanical and Aerospace Engineering, Monash University, Clayton, Victoria, Australia

[2]Department of Gastroenterology, The Alfred Hospital, Melbourne, Victoria, Australia

[Preprint]

**Abstract.**

*Background and Objectives*: Accurate and efficient prediction of soft tissue temperatures is essential to computer-assisted treatment systems for thermal ablation. It can be used to predict tissue temperatures and ablation volumes for personalised treatment planning and image-guided intervention. Numerically, it requires full nonlinear modelling of the coupled computational bioheat transfer and biomechanics, and efficient solution procedures; however, existing studies considered the bioheat analysis alone or the coupled linear analysis, without the fully coupled nonlinear analysis.

*Methods*: We present a coupled thermo-visco-hyperelastic finite element algorithm, based on finite-strain thermoelasticity and total Lagrangian explicit dynamics. It considers the coupled nonlinear analysis of (i) bioheat transfer under soft tissue deformations and (ii) soft tissue deformations due to thermal expansion/shrinkage. The presented method accounts for anisotropic, finite-strain, temperature-dependent, thermal, and viscoelastic behaviours of soft tissues, and it is implemented using GPU acceleration for real-time computation.

*Results*: The presented method can achieve thermo-visco-elastodynamic analysis of anisotropic soft tissues undergoing large deformations with high computational speeds in tetrahedral and hexahedral finite element meshes for surgical simulation of thermal ablation. We also demonstrate the translational benefits of the presented method for clinical applications using a simulation of thermal ablation in the liver.

*Conclusion*: The key advantage of the presented method is that it enables full nonlinear modelling of the anisotropic, finite-strain, temperature-dependent, thermal, and viscoelastic behaviours of soft tissues, instead of linear elastic, linear viscoelastic, and thermal-only modelling in the existing methods. It also provides high computational speeds for computer-assisted treatment systems towards enabling the operator to simulate thermal ablation accurately and visualise tissue temperatures and ablation zones immediately.

The source code is available at https://github.com/jinaojakezhang/FEDFEMBioheatExpan.

## 1. Introduction

Hyperthermia and thermal ablation treatments play an important role in the management of cancers following surgery, chemotherapy, and radiotherapy. Thermal ablation is a sub-category of hyperthermia in which local body tissues are heated until 50-100 ℃ over a few seconds or minutes to cause coagulation necrosis to damage and kill cancer cells [1]. The commonly used thermal ablative techniques include radiofrequency ablation (RFA), microwave ablation (MWA), laser ablation, and high-intensity focused ultrasound ablation (HIFU ablation). These techniques are typically minimally invasive (such as percutaneous RFA) or non-invasive (such as HIFU ablation), which are associated with fewer complications, shorter hospital stays, and less morbidity and mortality than those undergoing surgery.

Percutaneous thermal tumour ablation under imaging guidance is an important and widely accepted treatment option, often given with curative intent particularly in patients with early-stage hepatocellular carcinoma (HCC, primary liver cancer). The two most commonly employed methods are RFA and MWA, where RFA is currently the most widely recommended first-line ablation technique for those not suitable for surgery [2]. Conventional RFA involves a monopolar electrode that induces local heating using alternating current. Tissue heating results in coagulative necrosis, but temperatures decrease both with distance from the electrode and with the presence of blood flow. The latter phenomenon results in the so-called "heat sink effect" where adjacent large blood vessels cause tissue cooling, reducing the effectiveness of ablation [3].

MWA is a relatively recent and increasingly popular thermal ablative technique because of its ability to produce more rapid heating and higher maximum tissue temperatures. MWA has the advantages of producing wider and more predictable ablation volumes resulting in high complete ablation rates, and the ability to simultaneously treat multiple lesions [4] and potentially treat larger lesions more effectively [5]. Moreover, MWA requires less procedural time and is not subject to a heat-sink effect, which is known potential short-comings of RFA [6]. Rates of complications are low with both RFA (4.1%) and MWA (4.6%) [7, 8]. Still, despite the absence of convincing evidence, uptake of MWA is on the increase in many large academic centres in preference to RFA because of its ability to achieve a more rapid ablation, providing a workflow advantage particularly when performing multiple ablations. In the treatment of hepatic cancer, it was shown that the thermal ablation of solid, localised

Jinao.zhang@monash.edu; Sunita.chauhan@monash.edu

Department of Mechanical and Aerospace Engineering, Monash University, Wellington Road, Clayton, VIC 3800, Australia

small (≤ 3 $cm$) tumours was very effective, achieved complete tumour necrosis and provided similar survival outcomes to those undergoing liver resection [9].

To assist the operator of thermal ablation with better patient safety and greater ablation accuracy, computer-integrated treatment system is becoming increasingly important to provide *in silico* analysis of patient-specific soft tissue temperatures and thermal damage [10]. The aim of achieving a safe and effective ablation is to induce a desired level of thermal injury to the target tumour while avoiding unintended thermal damage to nearby healthy tissues. To this end, our objectives are to employ computational bioheat transfer and biomechanics to enable the operator to simulate thermal ablation accurately and visualise tissue temperatures and ablation zones on patient-specific anatomical models immediately. The challenges in achieving these are two-fold: (i) the mathematical models must be able to describe soft tissue thermo-mechanical responses accurately, and (ii) the numerical solution procedures must be formulated efficiently for fast computation.

The specific scenario we consider is towards real-time simulation of anisotropic thermo-visco-elastodynamic responses of soft tissues at finite-strains during thermal ablation. Due to body organ movements (such as respiratory motion) and tool-tissue interactions (such as percutaneous ablation), soft tissues are constantly in a state of motion consisted of dislocations and deformations; the former can be described by rigid-body displacements including translations and rotations, while the latter involves change in organ shapes that affect the distribution of thermal energy which is essential for computation of tissue temperatures. Furthermore, due to the variation of temperatures, the effects of thermal expansion [11] and shrinkage [12] induce thermal stresses, leading to thermally-induced deformations that also affect energy distributions. The thermally-induced deformations have also been evidenced in the studies of thermo-poroelastic modelling for fluid transport in tumours [13], thermal strain effects on low-density lipoprotein deposition in arterial walls [14], and thermo-diffusion effects [15]. Mathematically, these require full nonlinear modelling of the thermo-elastodynamic behaviour of soft tissues to account for (i) bioheat transfer under soft tissue deformations and (ii) soft tissue deformations due to thermal expansion/shrinkage. However, many of the existing studies were focused on thermal analysis alone without considering the mechanical behaviour [16-18]; therefore, these methods are incompatible and not suitable. Also, for such coupled nonlinear models, the need for efficient solution procedures for real-time computation on cost-effective and readily-available computing hardware for surgical simulation becomes more difficult [19].

Thermo-elastic analysis of soft tissues is not new to hyperthermia modelling; however, most of the existing methods are not suitable for surgical simulation. Key shortfalls of the existing methods stem from (i) expensive computational cost, (ii) solely steady-state analysis, and (iii) incompatibility with nonlinear thermo-elastic problems. Most of the reported methods [20, 21] were focused on sophisticated thermo-elastic models but failed to consider computational speed where solutions are often computationally expensive to obtain (e.g., Kröger *et al.* [22] reported a simulation of an 8 minutes RFA in three-dimensional (3D) space took about 6 hours). Some studies were focused on solely the steady-state thermo-elastic analysis [23], which ignored the transient thermo-elastic response of soft tissues in dynamics. Finally, some works considered only homogeneous and isotropic material properties with linear thermo-elasticity [21, 23] or linear thermo-visco-elasticity [24]. As a result, these works did not describe anisotropic viscoelastic properties of soft tissues and are incompatible with the nonlinear problem of soft tissues undergoing large deformations where geometric and material nonlinearities are involved.

To address the above issues, we develop a thermo-visco-hyperelastic total Lagrangian explicit dynamics finite element algorithm and utilise Graphics Processing Unit (GPU) acceleration for real-time finite-strain anisotropic thermo-visco-elastodynamic analysis of soft tissues. The following contributions are presented:

(i) we develop a thermo-visco-hyperelastic model that is fully nonlinear to describe the nonlinear characteristics of bioheat transfer in deformed soft tissues with thermal expansion/shrinkage (Section 3),
(ii) we develop efficient solution procedures for solving the coupled model in both tetrahedral and hexahedral computational grids (Section 4),
(iii) we develop a GPU implementation for real-time computation (Section 5), and
(iv) we demonstrate the translational benefits using a simulation of thermal ablation in the liver (Section 6.3).

As the outcome, the presented method can achieve thermo-visco-elastic analysis of anisotropic soft tissues undergoing large deformations in tetrahedral and hexahedral finite element meshes with high computational speeds for surgical simulation of thermal ablation.

The remainder of this work is organised as follows: Section 2 introduces the bioheat and biomechanics models, Sections 3, 4 and 5 present the proposed model, solution procedure, and GPU implementation, respectively. Algorithm verification, performance evaluation, and medical application are presented in Section 6. Discussions are presented in Section 7, and Section 8 concludes the present work.

## 2. Bioheat and biomechanics models

Computational thermo-visco-elastodynamic analysis of soft tissues requires the definition of bioheat and biomechanics models. We use the Pennes model for bioheat transfer and the visco-hyperelastic model for biomechanics analysis.

### 2.1 Bioheat transfer model

Various bioheat transfer models were reported in literature to characterise heat transfer in biological soft tissues, including Pennes model [25], Wulff model [26], Klinger model [27], Chen and Holmes model [28], porous-media model [29], discrete vasculature model [30], and dual-phase-lag model [31] (see review papers [1, 32]), among those the Pennes model is widely used [33-35] and has been compared against experimental data [36, 37] to provide reliable tissue temperature predictions in the absence of large blood vessels (diameters larger than $0.5\ mm$) [38]. Despite the development of other more sophisticated and rigorous bioheat models which may be seen as modifications from the classical Pennes model, they are usually obtained at the cost of increased numerical complexities and computational loads. Ge *et al.* [39] compared temperature profiles and histories between the Pennes model, dual-phase-lag model, and porous model, their results showed a lower temperature predicted by the porous model but nearly the same temperatures by the Pennes and the dual-phase-lag model. In addition, there exist controversies as to whether or not the dual-phase-lag conduction and any non-Fourier conduction is important for biological tissues [40] as some of the non-Fourier evidence has been called into question repeatedly [41, 42]. In light of this, we employ the Pennes model in the present work, which has proven to be remarkably effective for bioheat transfer analysis [43]; it is also computationally efficient and consumes less data storage space.

The Pennes bioheat model accounts for the thermal effects of anisotropic heat conduction in solid tissues, isotropic blood perfusion, metabolic heat generation, and regional heat sources. The governing equation of the transient Pennes model in 3D is given by [25]

$$^t\rho\ ^tc\frac{\partial\ ^tT(\mathbf{x})}{\partial t} = \nabla \cdot \left(\ ^tk\nabla\ ^tT(\mathbf{x})\right) -\ ^tw_b\ ^tc_b\left(\ ^tT(\mathbf{x}) -\ ^tT_a\right) +\ ^tQ_m +\ ^tQ_r \quad \forall \mathbf{x} \in \Omega \tag{1}$$

where $\rho$ is the tissue mass density, $c$ the specific heat capacity, $T(\mathbf{x})$ the temperature at a material point $\mathbf{x}(x,y,z)$ in the continuum $\Omega$, $t$ the time, $k$ the thermal conductivity, $w_b$ the blood perfusion rate, $c_b$ the blood specific heat capacity, $T_a$ the arterial blood temperature, $Q_m$ the metabolic heat generation rate, $Q_r$ the regional heat source rate, $\nabla \cdot$ the divergence operator, and $\nabla$ the gradient operator; a left superscript $^t\bullet$ denotes the system configuration in which a quantity occurs ($t$ denotes the current time system configuration); the material properties are, in general, temperature-dependent [44, 45] and can be directly incorporated for tissue temperature analysis.

Soft tissue temperatures must be determined from the deformed configuration due to dynamic motions of soft tissues; however, the classical Pennes model is based on only the static non-moving state [46]. To address this, we apply a mapping function to transform the unknown current (deformed) configuration to the known reference (undeformed) configuration for bioheat transfer analysis, which will be demonstrated in Section 3.1.

## 2.2 Visco-hyperelastic biomechanics model

Soft tissues undergo large deformations under external forces where geometric and material nonlinearities are involved, and they exhibit time-dependent viscous and elastic behaviours [47]. The widely used linear elastic and linear viscoelastic models are based on the assumptions of infinitesimally small deformations with the linear stress-strain relationship, but they are not compatible with large deformations of soft tissues. For fully compatible nonlinear modelling, we use the visco-hyperelastic biomechanics model based on finite elasticity for which nonlinear continuum mechanics is the fundamental basis. By referring to the reference system configuration, the variation of strain at a material point is expressed by the Green-Saint Venant strain tensor $^t_0\mathbf{E} = \frac{1}{2}(^t_0\mathbf{C} - \mathbf{I})$ where $^t_0\mathbf{C}$ is the right Cauchy-Green tensor and $\mathbf{I}$ is the identity matrix of the second rank. The corresponding strain energy density is expressed by $^t_0\Psi$ which is typically a function of the strain invariants of $^t_0\mathbf{C}$. Accordingly, the energy conjugated stress measure is the second Piola-Kirchhoff stress tensor $^t_0\mathbf{S} = \frac{\partial\ ^t_0\Psi(^t_0\mathbf{C})}{\partial\ ^t_0\mathbf{E}}$. Mechanical behaviours are accommodated in $^t_0\Psi$. For anisotropic soft tissues exhibiting directional-dependent behaviours, $^t_0\Psi$ can be modified by employing unit vectors to describe local fibre directions [48]. For viscoelastic behaviours, $^t_0\Psi$ can be modified by employing a time-dependent $^t_0\widetilde{\Psi}$ such as $^t_0\widetilde{\Psi} = \int_0^t \varphi(t-t')\frac{\partial\ ^t_0\Psi}{\partial t'}\,dt'$ in a convolution integral [49] where $\varphi(t-t') = \varphi_\infty + \sum_{i=1}^N \varphi_i e^{(t'-t)/\tau_i}$ is the relaxation function of a generalised Maxwell model [50], and $\varphi_\infty$, $\varphi_i$, and $\tau_i$ are positive constants.

By referring to the reference configuration, the equation of motion is given by [51]

$$\nabla \cdot\ ^t_0\mathbf{P} +\ ^t_0\mathbf{B} =\ ^0\rho\ ^t\dot{\mathbf{V}} \tag{2}$$

where $\mathbf{P}$ is the first Piola-Kirchhoff stress tensor, $\mathbf{B}$ the reference body force, and $\dot{\mathbf{V}}$ the material acceleration field; a left subscript $_0\bullet$ denotes the configuration with respect to which the quantity is measured (0 denotes the reference configuration).

The above balance principle is valid for any arbitrary parts of a continuum, but the stress measure is expressed for elastodynamic analysis. For thermo-elastodynamic analysis, the constitutive expression for the stress is obtained based on the Helmholtz free energy which is a function of the finite-strain and temperature and by exploring the conservation of energy and the second law of thermodynamics [52]. To obtain this stress expression, we apply a multiplicative decomposition [52] to include the elastic and thermal deformations for the total stress, which will be demonstrated in Section 3.2.

## 3. Coupling of bioheat and biomechanics models

A two-way interaction of bioheat and biomechanics models is identified: (i) due to soft tissue deformations, transient tissue temperatures must be computed based on the deformed configuration, and (ii) thermal expansion/shrinkage also induces soft tissue deformations. Both (i) and (ii) affect the prediction of thermal ablation outcomes and require the coupled thermo-elastodynamic modelling.

Fig. 1 illustrates the notations used in the proposed coupled thermo-elastodynamic analysis. Using the Lagrangian description, soft tissues undergo deformations $^0\Omega \to {}^t\Omega$ described by the deformation gradient ${}_0^t\mathbf{F}$. An intermediate system configuration $^i\Omega$ is introduced, decomposing the total deformation gradient ${}_0^t\mathbf{F}$ into ${}_0^i\mathbf{F}(\,^0\Omega \to {}^i\Omega)$ and ${}_i^t\mathbf{F}(\,^i\Omega \to {}^t\Omega)$. We compute (i) bioheat transfer under deformations by transformation of the system configuration $^t\Omega \to {}^0\Omega$ (Section 3.1) and (ii) soft tissue deformations due to thermal expansion/shrinkage by multiplicative decomposition ${}_0^t\mathbf{F} = {}_i^t\mathbf{F}\,{}_0^i\mathbf{F}$ (Section 3.2).

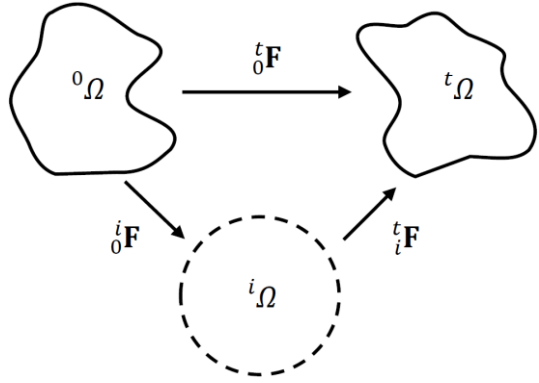

Fig. 1 Reference $^0\Omega$, intermediate $^i\Omega$, and current $^t\Omega$ configurations, and the deformations described by ${}_0^t\mathbf{F}(\,^0\Omega \to {}^t\Omega)$, ${}_0^i\mathbf{F}(\,^0\Omega \to {}^i\Omega)$, and ${}_i^t\mathbf{F}(\,^i\Omega \to {}^t\Omega)$.

### 3.1 Mathematical modelling of bioheat transfer under soft tissue deformations

We employ a transformation of system configurations to achieve bioheat computation on deformed soft tissues [53]. The mapping function for deformations $^0\Omega \to {}^t\Omega$ can be defined by

$$^t\xi : {}^0\Omega \times \mathbb{R}_0^+ \to {}^t\Omega, \qquad {}^t\mathbf{x} = {}^t\xi(\,^0\mathbf{x}) = \xi(\,^0\mathbf{x}, t) \tag{3}$$

where the mapping function $\xi$ is continuous in time, continuously differentiable, globally invertible, and orientation-preserving at all times.

Using the current (deformed) configuration $^t\Omega$, the Pennes model can be integrated over the volume to get

$$\int_{^t\Omega} \left(\,^t\rho\,{}^tc\frac{\partial\,^tT(\mathbf{x})}{\partial t}\right) d\,^t\Omega = \int_{^t\Omega} \left(\,^t\nabla \cdot (\,^tk\,{}^t\nabla\,{}^tT(\mathbf{x})) - {}^tw_b\,{}^tc_b(\,^tT(\mathbf{x}) - {}^tT_a) + {}^tQ_m + {}^tQ_r\right) d\,^t\Omega \tag{4}$$

By performing a change of variables, it can be expressed in terms of the reference configuration $^0\Omega$ as

$$\int_{^t\Omega} (\ldots)\,d\,^t\Omega = \int_{^0\Omega} (\ldots)\det\left(\mathcal{L}\xi(\,^0\mathbf{x},t)\right) d\,^0\Omega = \int_{^0\Omega} (\ldots)\det\left(\frac{\partial\,^t\mathbf{x}}{\partial\,^0\mathbf{x}}\right) d\,^0\Omega = \int_{^0\Omega} (\ldots)\det({}_0^t\mathbf{F})\,d\,^0\Omega \tag{5}$$

where $\mathcal{L}$ is the differential operator.

The divergence and gradient operators ($^t\nabla\cdot$ and $^t\nabla$) are expressed in terms of $^t\Omega$ which is unknown. By applying the chain rule, they can be expressed in terms of the known reference configuration $^0\Omega$ as

$$\int_{^t\Omega} {}^t\nabla \cdot (\,^tk\,{}^t\nabla\,{}^tT(\mathbf{x}))\,d\,^t\Omega = \int_{^0\Omega} \left(\,^0\nabla_0^t\mathbf{F}^{-1} \cdot (\,^tk\,{}^0\nabla_0^t\mathbf{F}^{-1}\,{}^tT(\mathbf{x}))\right)\det({}_0^t\mathbf{F})\,d\,^0\Omega \tag{6}$$

The above formulation allows for the computation of bioheat transfer on deformed configurations of soft tissues, compared to the classical Pennes model which was computed based on only the static non-moving state [46]. From the point of view of numerical efficiency, the above formulation is based on only ${}_0^t\mathbf{F}(\,^0\Omega \to {}^t\Omega)$, $^tk$, and $^tT(\mathbf{x})$, whereas the other variables are defined over the reference configuration $^0\Omega$ that can be precomputed. Mathematically, it accounts for nonlinear characteristics of bioheat transfer at finite strains.

### 3.2 Mathematical modelling of soft tissue deformations due to thermal expansion/shrinkage

We apply a multiplicative decomposition to build the stress expression for thermo-elastodynamic analysis. Since the deformations caused by thermal expansion and shrinkage are mathematically equivalent, we will use thermal expansion for the volume change of soft tissues in the remainder of the present work. Prior to the onset of thermal expansion, we consider the tissue is in a physiological state described by the stress-free intermediate configuration ($^i\Omega$ in Fig. 1), which is a hypothetical configuration obtained by isothermal elastic destressing of the deformed configuration $^t\Omega$ to zero stress [54]. With this intermediate configuration, the total deformation gradient $_0^t\mathbf{F}$ for $^0\Omega \rightarrow \,^t\Omega$ can be decomposed into a thermal part $_0^i\mathbf{F}^{ther}$ and an elastic part $_i^t\mathbf{F}^{elas}$ via a local multiplicative decomposition [52] expressed by

$$_0^t\mathbf{F} = \,_i^t\mathbf{F}^{elas}\,_0^i\mathbf{F}^{ther} \tag{7}$$

To meet the requirement for zero stress under zero strain, the constitutive expression must be written in terms of the elastic deformation gradient $_i^t\mathbf{F}^{elas}$, yielding

$$_i^t\mathbf{F}^{elas} = \,_0^t\mathbf{F}\,_0^i\mathbf{F}^{ther^{-1}} \tag{8}$$

The second Piola-Kirchhoff stress $_i^t\mathbf{S}$ associated with the stress-free intermediate configuration $^i\Omega$ can then be expressed by

$$_i^t\mathbf{S} = \frac{\partial\,^t\Psi(_i^t\mathbf{C})}{\partial\,_i^t\mathbf{E}} = 2\frac{\partial\,^t\Psi(_i^t\mathbf{C})}{\partial\,_i^t\mathbf{C}} = \,_i^t\mathbf{S}(_i^t\mathbf{F}^{elas}) = \,_i^t\mathbf{S}\left(_0^t\mathbf{F}\,_0^i\mathbf{F}^{ther^{-1}}\right) \tag{9}$$

By pulling back the stress to the reference configuration $^0\Omega$, the total stress $_0^t\mathbf{S}$ can be expressed by

$$_0^t\mathbf{S} = \det\bigl(_0^i\mathbf{F}^{ther}\bigr)\,_0^i\mathbf{F}^{ther^{-1}}\,_i^t\mathbf{S}\left(_0^t\mathbf{F}\,_0^i\mathbf{F}^{ther^{-1}}\right)\,_0^i\mathbf{F}^{ther^{-T}} \tag{10}$$

The above formulation is based on the deformation gradients $_0^t\mathbf{F}$ and $_0^i\mathbf{F}^{ther}$, which enables thermal expansion to be included entirely in the material model.

The thermal deformation gradient $_0^i\mathbf{F}^{ther}$ can be specified depending on the type of material anisotropy. For an orthotropic thermal expansion with the principal axes parallel to unit vectors $^0\mathbf{m}$, $^0\mathbf{n}$ and $^0\mathbf{m} \times \,^0\mathbf{n}$ in $^0\Omega$, $_0^i\mathbf{F}^{ther}$ can be specified by [54]

$$_0^i\mathbf{F}^{ther} = \lambda_i\mathbf{I} + (\lambda_m - \lambda_i)\,^0\mathbf{m} \otimes \,^0\mathbf{m} + (\lambda_n - \lambda_i)\,^0\mathbf{n} \otimes \,^0\mathbf{n} \tag{11}$$

where $\lambda_i$, $\lambda_m$ and $\lambda_n$ are the stretch ratios in the orthogonal directions $^0\mathbf{m} \times \,^0\mathbf{n}$, $^0\mathbf{m}$ and $^0\mathbf{n}$, respectively; the above expression can be modified for transversely isotropic thermal expansion $_0^i\mathbf{F}^{ther} = \lambda_i\mathbf{I} + (\lambda_m - \lambda_i)\,^0\mathbf{m} \otimes \,^0\mathbf{m}$ and isotropic thermal expansion $_0^i\mathbf{F}^{ther} = \lambda_i\mathbf{I}$. The thermal stretch ratio $\lambda$ is related to the thermal expansion coefficient $\alpha$ by $\lambda \cong 1 + \alpha\bigl(\,^tT(\mathbf{x}) - \,^0T(\mathbf{x})\bigr)$ [52].

## 4. Numerical solution procedures

We develop a thermo-visco-hyperelastic total Lagrangian explicit dynamics finite element algorithm for solutions of the coupled bioheat and biomechanics analysis.

### 4.1 Thermo-visco-hyperelastic total Lagrangian explicit dynamics finite element algorithm

To achieve fast and accurate numerical solutions, we develop solution procedures based on the (i) fast explicit dynamics finite element algorithm [55] for transient bioheat computation [56] under soft tissue deformations [53] and the (ii) total Lagrangian explicit dynamics finite element algorithm [57] for biomechanics computation of large-strain soft tissue deformations. The proposed formulation has not been evidenced in previous works.

We employ the total Lagrangian formulation for computational biomechanics for its numerical efficiency without error accumulation. Displacement-based finite element formulations use the updated Lagrangian formulation or the total Lagrangian formulation to describe the frame of reference [58]. The updated Lagrangian obtains variable values at the current configuration based on the end of the previous time step via an incremental strain-description, which requires a re-calculation of spatial derivatives at each time step. In contrast, the total Lagrangian considers all variables referred to the reference configuration so that the strains lead to correct results after a load cycle without error accumulation [59], which is not the case in the incremental strain-description. More importantly, it allows for the spatial derivatives to be precomputed and stored [57]. It was shown that the total Lagrangian took 2.1 $ms$ for a time step to obtain the solution for an ellipsoid indentation, which was about five times faster compared to 10.6 $ms$ consumed by the updated Lagrangian [60].

We employ explicit finite element method (FEM) and explicit time integration for computational bioheat transfer and biomechanics for numerical efficiency and parallelisation. Explicit FEM [61] is a simplified form of FEM that is often used in surgical simulation. The internal and external loads and mass are lumped to the nodes, leading to block diagonal mass and damping matrices to enable computation to be performed at the element level without the need for assembling the stiffness matrix for the entire model [57]. The explicit FEM can be integrated in time either explicitly or implicitly for dynamics, and we use explicit time integration to allow for variable values in future states to be computed directly based on the current state without the need for stiffness matrix inversion which is required in the implicit integration. It was shown that the implicit integration consumed computation time that was at least one order of magnitude larger than that by the explicit counterpart [62]. The resultant explicit dynamics leads to an explicit formulation for unknown state variables that is well suited for parallel computation. The global system of equations can be split into independent equations for individual nodes, allowing each nodal computation to be assigned to a parallel task to perform calculations independently. Along with computationally efficient low-order finite elements, the above formulations lead to very efficient computation at run-time.

### 4.2 Formulation for bioheat transfer under soft tissue deformations

Using the first-order explicit time integration and the lumped (diagonalised) mass approximation while ensuring the conservation of mass, the discretized matrix equation of the Pennes model can be written as

$$^t\mathbf{C}^{diag}\left(\frac{^{t+\Delta t}\mathbf{T}(\mathbf{x}) - {}^t\mathbf{T}(\mathbf{x})}{\Delta t}\right) = {}^t\mathbf{K}\ {}^t\mathbf{T}(\mathbf{x}) - {}^t\mathbf{K}_b^{diag}\ {}^t\mathbf{T}(\mathbf{x}) + {}^t\mathbf{Q}_b + {}^t\mathbf{Q}_m + {}^t\mathbf{Q}_r \tag{12}$$

where $\mathbf{C}^{diag}$ is the diagonalised thermal mass (mass and specific heat capacity) matrix, $\mathbf{T}(\mathbf{x})$ the vector of nodal temperatures, $\Delta t$ the time step, $\mathbf{K}$ the thermal stiffness (conduction) matrix, $\mathbf{K}_b^{diag}$ the diagonalised thermal stiffness (blood perfusion) matrix, $\mathbf{Q}_b$, $\mathbf{Q}_m$, and $\mathbf{Q}_r$ the vectors of heat flow of blood perfusion at arterial temperature, metabolic heat generation, and regional heat sources, respectively. The equation can be further rearranged into

$$^{t+\Delta t}\mathbf{T}(\mathbf{x}) = {}^t\mathbf{T}(\mathbf{x}) + \Delta t\ {}^t\mathbf{C}^{diag^{-1}}\left(\sum_e {}_0^t\mathbf{f}_e^{ther} - {}^t\mathbf{K}_b^{diag}\ {}^t\mathbf{T}(\mathbf{x}) + {}^t\mathbf{Q}_b + {}^t\mathbf{Q}_m + {}^t\mathbf{Q}_r\right) \tag{13}$$

where

$$\sum_e {}_0^t\mathbf{f}_e^{ther} = {}^t\mathbf{K}\ {}^t\mathbf{T}(\mathbf{x}) = {}_0^t\mathbf{f}^{ther} \tag{14}$$

where $\mathbf{f}_e^{ther}$ is the vector of thermal load components due to heat conduction in an element $e$, and $\mathbf{f}^{ther}$ is the vector of global nodal thermal loads. ${}_0^t\mathbf{f}_e^{ther}$ is computed by

$$_0^t\mathbf{f}_e^{ther} = \int_{^tV_e} \mathcal{L}\ {}^t\mathbf{h_x}^T\ {}^t\mathbf{D}_e \mathcal{L}\ {}^t\mathbf{h_x}\ d\ {}^tV\ {}^t\mathbf{T}(\mathbf{x}) \tag{15}$$

where $\mathcal{L}\mathbf{h_x}$ is the matrix of element shape function spatial derivatives, $\mathbf{D}_e$ the element thermal conductivity matrix, and $V_e$ the element volume.

By incorporating the proposed bioheat formulation under soft tissue deformations (Eq. (6)), the above equation at the current configuration $^t\Omega$ can be written in terms of the reference configuration $^0\Omega$ as

$$_0^t\mathbf{f}_e^{ther} = \int_{^0V_e} \left(\mathcal{L}\ {}^0\mathbf{h_x}{}_0^t\mathbf{F}^{-1}\right)^T\ {}^t\mathbf{D}_e\left(\mathcal{L}\ {}^0\mathbf{h_x}{}_0^t\mathbf{F}^{-1}\right)\det({}_0^t\mathbf{F})\ d\ {}^0V\ {}^t\mathbf{T}(\mathbf{x}) \tag{16}$$

where ${}_0^t\mathbf{F}$ can be computed from the element nodal displacements matrix $\mathbf{u}_e$ by

$$_0^t\mathbf{F} = \left({}^t\mathbf{u}_e\right)^T \mathcal{L}\ {}^0\mathbf{h_x} + \mathbf{I} \tag{17}$$

For fast computation, the computationally efficient 3D low-order finite elements such as the eight-node reduced-integrated hexahedral elements and the four-node linear tetrahedral elements are used, leading to the following computations at the element level:

(i) eight-node reduced-integrated hexahedral element:

$$_0^t\mathbf{f}_e^{ther} = 8\det({}^0\mathbf{J})\left(\mathcal{L}\ {}^0\mathbf{h_x}{}_0^t\mathbf{F}^{-1}\right)^T\ {}^t\mathbf{D}_e\mathcal{L}\ {}^0\mathbf{h_x}{}_0^t\mathbf{F}^{-1}\det({}_0^t\mathbf{F})\ {}^t\mathbf{T}(\mathbf{x}) \tag{18}$$

(ii) four-node linear tetrahedral element:

$$_0^t\mathbf{f}_e^{ther} = {}^0V\left(\mathcal{L}\ {}^0\mathbf{h_x}{}_0^t\mathbf{F}^{-1}\right)^T\ {}^t\mathbf{D}_e\mathcal{L}\ {}^0\mathbf{h_x}{}_0^t\mathbf{F}^{-1}\det({}_0^t\mathbf{F})\ {}^t\mathbf{T}(\mathbf{x}) \tag{19}$$

where $\mathbf{J}$ is the element Jacobian matrix.

At each time step, the temperature $^{t+\Delta t}T_{(i)}(\mathbf{x})$ at a node $i$ can be obtained by

$$^{t+\Delta t}T_{(i)}(\mathbf{x}) = {}^tT_{(i)}(\mathbf{x}) + \Delta t\ {}^tC_{(i)}^{diag^{-1}}\left({}_0^tf_{(i)}^{ther} - {}^tK_{b(i)}^{diag}\ {}^tT_{(i)}(\mathbf{x}) + {}^tQ_{b(i)} + {}^tQ_{m(i)} + {}^tQ_{r(i)}\right) \tag{20}$$

The above formulation allows for the computation of bioheat transfer with respect to the deformed configuration of soft tissues. Eq. (20) states an explicit formulation for the unknown temperature $^{t+\Delta t}T_{(i)}(\mathbf{x})$ which can be obtained based on the variable values at $t$ only; hence, the global system of equations can be split into independent equations for parallel computation. Furthermore, there is no need for iterations during any process of the algorithm. Temperature-dependent thermal properties and nonlinear thermal boundary conditions can also be directly incorporated.

### 4.3 Formulation for soft tissue deformations due to thermal expansion

The discretized matrix equation of the mechanics of motion can be written as

$$^{t}\mathbf{M}^{diag}\frac{\partial^{2}\,{}^{t}\mathbf{U}(\mathbf{x})}{\partial t^{2}}+{}^{t}\mathbf{D}^{diag}\frac{\partial\,{}^{t}\mathbf{U}(\mathbf{x})}{\partial t}+{}^{t}\mathbf{K}(\mathbf{U})\,{}^{t}\mathbf{U}(\mathbf{x})={}^{t}\mathbf{R} \quad (21)$$

where $\mathbf{M}^{diag}$ is the diagonalised mass matrix, $\mathbf{D}^{diag}=\gamma\mathbf{M}^{diag}$ the mass-proportional damping (a special case of Rayleigh damping, $\gamma$ the damping coefficient), $\mathbf{K}(\mathbf{U})$ the stiffness matrix, $\mathbf{U}(\mathbf{x})$ the vector of nodal displacements, and $\mathbf{R}$ the vector of externally applied forces.

The displacements $^{t+\Delta t}\mathbf{U}(\mathbf{x})$ at the next time step can be computed based on the explicit central-difference scheme [63], yielding

$$^{t+\Delta t}\mathbf{U}(\mathbf{x})=\frac{{}^{t}\mathbf{R}-\sum_{e}{}_{0}^{t}\mathbf{f}_{e}^{ther-elas}+\frac{2\,{}^{t}\mathbf{M}^{diag}}{\Delta t^{2}}\,{}^{t}\mathbf{U}(\mathbf{x})+\left(\frac{{}^{t}\mathbf{D}^{diag}}{2\Delta t}-\frac{{}^{t}\mathbf{M}^{diag}}{\Delta t^{2}}\right){}^{t-\Delta t}\mathbf{U}(\mathbf{x})}{\frac{{}^{t}\mathbf{D}^{diag}}{2\Delta t}+\frac{{}^{t}\mathbf{M}^{diag}}{\Delta t^{2}}} \quad (22)$$

where

$$\sum_{e}{}_{0}^{t}\mathbf{f}_{e}^{ther-elas}={}^{t}\mathbf{K}(\mathbf{U})\,{}^{t}\mathbf{U}(\mathbf{x})={}_{0}^{t}\mathbf{f}^{ther-elas} \quad (23)$$

where $\mathbf{f}_{e}^{ther-elas}$ is the vector of nodal forces due to thermal and elastic stresses in an element $e$, and $\mathbf{f}^{ther-elas}$ is the vector of global nodal forces. ${}_{0}^{t}\mathbf{f}_{e}^{ther-elas}$ can be computed by

$$_{0}^{t}\mathbf{f}_{e}^{ther-elas}=\int_{{}^{0}V_{e}}{}_{0}^{t}\mathbf{F}{}_{0}^{t}\mathbf{S}\mathcal{L}\,{}^{0}\mathbf{h}_{\mathbf{x}}\,d\,{}^{0}V \quad (24)$$

where ${}_{0}^{t}\mathbf{S}$ is the total stress to account for the effect of thermal expansion given by Eq. (10).

The corresponding low-order finite element formulations are:

(i) eight-node reduced-integrated hexahedral element:

$$_{0}^{t}\mathbf{f}_{e}^{ther-elas}=8\det({}^{0}\mathbf{J}){}_{0}^{t}\mathbf{F}\left(\det({}_{0}^{i}\mathbf{F}^{ther}){}_{0}^{i}\mathbf{F}^{ther-1}{}_{i}^{t}\mathbf{S}\left({}_{0}^{t}\mathbf{F}{}_{0}^{i}\mathbf{F}^{ther-1}\right){}_{0}^{i}\mathbf{F}^{ther-T}\right)\mathcal{L}\,{}^{0}\mathbf{h}_{\mathbf{x}} \quad (25)$$

(ii) four-node linear tetrahedral element:

$$_{0}^{t}\mathbf{f}_{e}^{ther-elas}={}^{0}V\,{}_{0}^{t}\mathbf{F}\left(\det({}_{0}^{i}\mathbf{F}^{ther}){}_{0}^{i}\mathbf{F}^{ther-1}{}_{i}^{t}\mathbf{S}\left({}_{0}^{t}\mathbf{F}{}_{0}^{i}\mathbf{F}^{ther-1}\right){}_{0}^{i}\mathbf{F}^{ther-T}\right)\mathcal{L}\,{}^{0}\mathbf{h}_{\mathbf{x}} \quad (26)$$

At each time step, the displacement component $^{t+\Delta t}u_{x(i)}$ of $^{t+\Delta t}\mathbf{u}_{(i)}\left({}^{t+\Delta t}u_{x(i)},{}^{t+\Delta t}u_{y(i)},{}^{t+\Delta t}u_{z(i)}\right)$ at a node $i$ can be obtained by

$$^{t+\Delta t}u_{x(i)}=\frac{{}^{t}R_{x(i)}-{}_{0}^{t}f_{x(i)}^{ther-elas}+\frac{2\,{}^{t}M_{x(i)}^{diag}}{\Delta t^{2}}\,{}^{t}u_{x(i)}+\left(\frac{{}^{t}D_{x(i)}^{diag}}{2\Delta t}-\frac{{}^{t}M_{x(i)}^{diag}}{\Delta t^{2}}\right){}^{t-\Delta t}u_{x(i)}}{\frac{{}^{t}D_{x(i)}^{diag}}{2\Delta t}+\frac{{}^{t}M_{x(i)}^{diag}}{\Delta t^{2}}} \quad (27)$$

For viscoelastic modelling where the time-dependent ${}_{0}^{t}\widetilde{\Psi}=\int_{0}^{t}\varphi(t-t')\frac{\partial\,{}_{0}^{t}\Psi}{\partial t'}\,dt'$ is used (Section 2.2), the corresponding time-dependent stress is expressed by ${}_{0}^{t}\widetilde{\mathbf{S}}=\int_{0}^{t}\varphi(t-t')\frac{\partial\,{}_{0}^{t}\mathbf{S}}{\partial t'}\,dt'$, and it can be computed by ${}_{0}^{t}\widetilde{\mathbf{S}}={}_{0}^{t}\mathbf{S}-\sum_{i=1}^{N}{}_{0}^{t}\boldsymbol{\vartheta}_{i}$ [64] where ${}_{0}^{t}\boldsymbol{\vartheta}_{i}=\frac{\Delta t\varphi_{i}}{\Delta t+\tau_{i}}{}_{0}^{t}\mathbf{S}+\frac{\tau_{i}}{\Delta t+\tau_{i}}{}_{0}^{t-\Delta t}\boldsymbol{\vartheta}_{i}$. The corresponding thermo-visco-elastic nodal forces ${}_{0}^{t}\mathbf{f}_{e}^{ther-visco-elas}$ can be obtained by substituting the time-dependent ${}_{0}^{t}\widetilde{\mathbf{S}}$ for ${}_{0}^{t}\mathbf{S}$ in Eq. (24), yielding

$$_{0}^{t}\mathbf{f}_{e}^{ther-visco-elas}=\int_{{}^{0}V_{e}}{}_{0}^{t}\mathbf{F}{}_{0}^{t}\widetilde{\mathbf{S}}\mathcal{L}\,{}^{0}\mathbf{h}_{\mathbf{x}}\,d\,{}^{0}V \quad (28)$$

which is used to replace ${}_{0}^{t}f_{x(i)}^{ther-elas}$ by ${}_{0}^{t}f_{x(i)}^{ther-visco-elas}$ in Eq. (27) for viscoelastic modelling.

The above formulation enables the time-dependent effect to be included entirely in the thermo-elastic stress for thermo-visco-elastic constitutive analysis. The constitutive material law, temperature-dependent mechanical properties and nonlinear mechanical boundary conditions can be directly incorporated. Fig. 2 illustrates an implementation of the proposed numerical solution procedures.

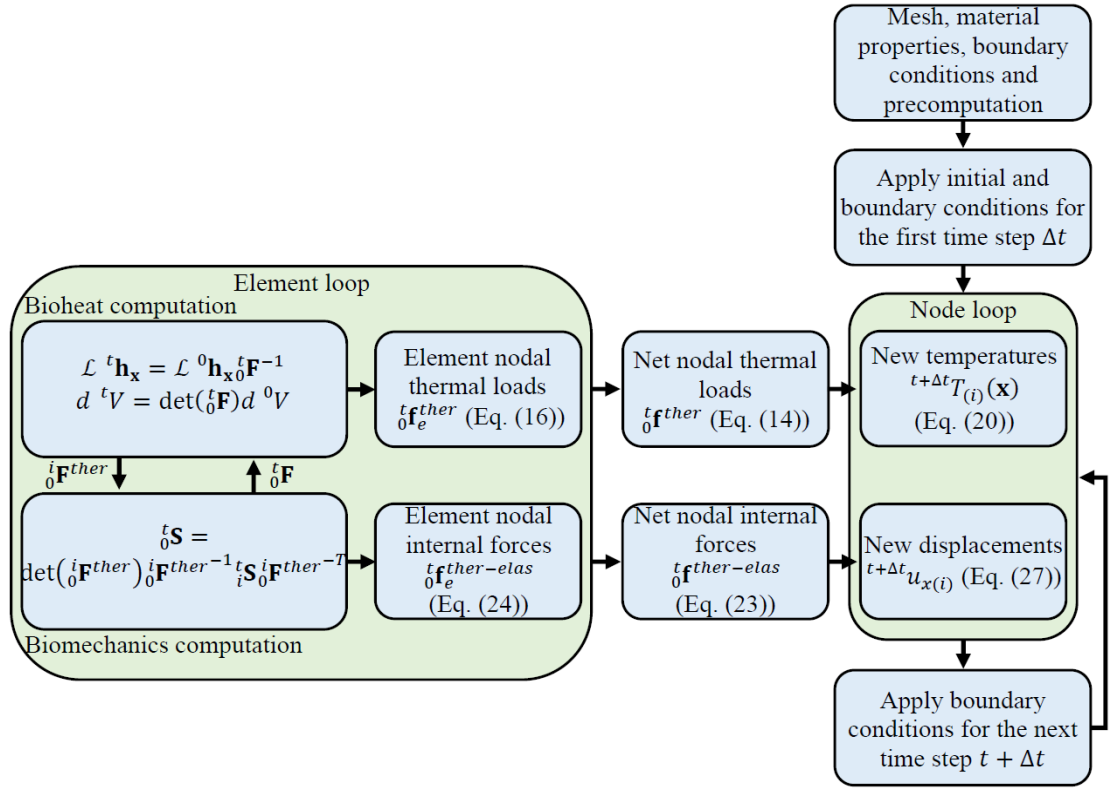

Fig. 2. Key components of the proposed numerical solution procedures.

## 5. GPU implementation

Having established the formulations for bioheat transfer under soft tissue deformations (Section 4.2) and soft tissue deformations due to thermal expansion (Section 4.3) for tetrahedral and hexahedral computational grids, the presented method is implemented using GPU parallel computation for real-time surgical simulation. The GPU implementation consists of a host (CPU) and a device (GPU) code for host-device interaction and device parallel computation. The host implementation is responsible for precomputing and storing the constant matrices and simulation parameters and for invoking device methods to interact with GPUs for memory allocation, texture binding, data copy from/to device, and kernel launching. As mentioned previously, many simulation parameters can be precomputed owing to the total Lagrangian formulation, such as the spatial derivatives, initial element volumes, initial Jacobian, and mass and damping matrices. The host code is written in the C++ programming language in Visual Studio 2017.

The device implementation is responsible for computing new internal thermal loads, temperatures, internal forces and displacements for the temperature and displacement fields. The nodal loads, such as the thermal loads and internal forces, are computed by an "element" kernel for tetrahedron/hexahedron calculations. The nodal temperatures and displacements are computed by Eqs. (20) and (27), respectively, at each time step for time-stepping, and they are computed by a "node" kernel for node calculation. A simulation time step is achieved by launching the "element" kernel across $n_e$ threads and the "node" kernel across $n_n$ threads where $n_e$ and $n_n$ denote the numbers of finite elements and nodes in the organ models, respectively. The GPU-accelerated computation is achieved via a time-loop of such time-step computation. The device code is written using the NVIDIA CUDA programming API version 10.2.

## 6. Numerical results

Numerical evaluations are conducted to assess the validity of the presented methodology for simulating large strain thermo-elastodynamics of soft tissues for surgical simulation, and assessments on CPU and GPU computational performance are presented. The translational benefits for a clinically relevant application are demonstrated using a simulation of thermal ablation in the liver.

### 6.1 Algorithm verification

Although living tissues need to be used for validation of the proposed method, experiments with living tissues raise ethical issues with technical difficulties. In the present work, a 3D numerical example with thermal and mechanical properties similar to those of soft tissues is used for algorithm verification, which offers a controlled environment for a quantitative assessment on numerical accuracy.

Fig. 3 illustrates the geometry and simulation settings of the employed model, and Table. 1 presents the values of the geometry, material properties and simulation time. The prescribed displacement $u_z = 0.02\ m$ in Fig. 3 corresponds to 40% extension compared to the height ($0.05\ m$) of the model for large deformations. The right face of the model was assumed to be fixed in position during the simulation. Adiabatic boundary condition was applied to the exterior of the model. The established nonlinear procedures, 'Dynamic, Temp-disp, Explicit', from commercial finite element analysis package, ABAQUS/CAE 2018 (2017_11_08-04.21.41 127140), were used to produce reference solutions under the same conditions for comparison. To quantitatively evaluate the validity of the presented method, normalised relative errors $NRE_T = \left|\frac{T_i^{ABAQUS}-T_i^{Proposed}}{T_{max}^{ABAQUS}-T_{min}^{ABAQUS}}\right|$ and $NRE_u = \left|\frac{u_i^{ABAQUS}-u_i^{Proposed}}{u_{max}^{ABAQUS}-u_{min}^{ABAQUS}}\right|$ and total error indicators $e_T = \sqrt{\frac{\sum_{i=1}^{n}\left(T_i^{ABAQUS}-T_i^{Proposed}\right)^2}{\sum_{i=1}^{n}\left(T_i^{ABAQUS}\right)^2}}$ and $e_u = \sqrt{\frac{\sum_{i=1}^{n}\left(u_i^{ABAQUS}-u_i^{Proposed}\right)^2}{\sum_{i=1}^{n}\left(u_i^{ABAQUS}\right)^2}}$ were used; $u_i$ included the three components $u_{i,x}, u_{i,y}$ and $u_{i,z}$.

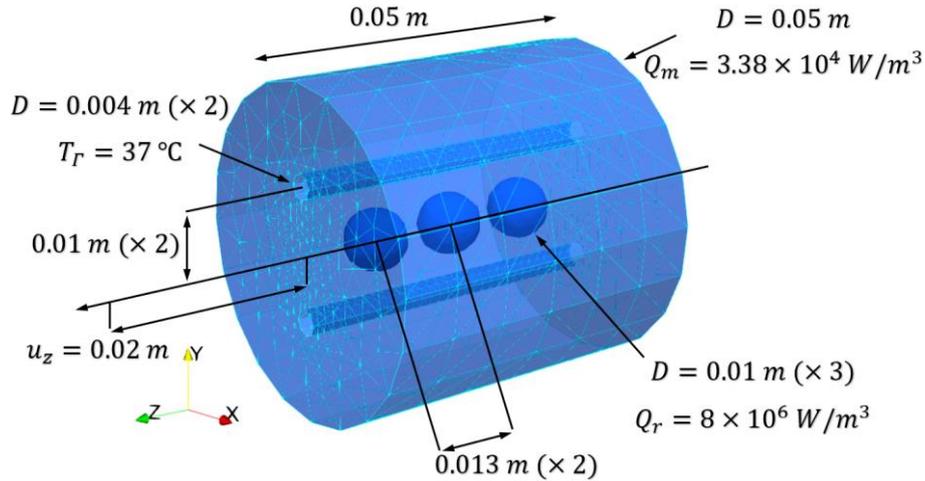

Fig. 3. Model geometry and simulation settings for algorithm verification.

Table. 1 Simulation settings of the model in Fig. 3.

| Geometry | Values | Unit | Description |
| --- | --- | --- | --- |
| Nodes | 4489 | | 17956 degrees of freedom (DOFs, $x, y, z, T$ per node) |
| Elements | 24290 | | Four-node linear tetrahedrons |
| Material properties | | | |
| Density $\rho$ | 1060 | $[kg/m^3]$ | [21] |
| Specific heat capacity $c$ | 3700 | $[J/(kg \cdot °C)]$ | [21] |
| Thermal conductivity $k$ | 0.518 | $[W/(m \cdot °C)]$ | [21] |
| Thermal expansion $\alpha$ | 0.1 | $[1/°C]$ | |
| Isotropic neo-Hookean hyperelastic model | | | $\Psi = \frac{\mu}{2}(\bar{I}_1 - 3) + \frac{\kappa}{2}(J-1)^2$ |
| Shear modulus $\mu$ | 1190.476 | $[Pa]$ | Corresponding to Young's modulus 3500 Pa and Poisson's ratio 0.47 [53] |
| Bulk modulus $\kappa$ | 19444.444 | $[Pa]$ | |
| Simulation time | $5\ s$ | Time step | $0.00008\ s$ (62500 steps) |
| Note: the anisotropic thermal conductivity, anisotropic hyperelastic material models, and viscoelasticity have been previously verified in Refs. [55] and [64]. | | | |

Fig. 4 presents a comparison between four cases of the numerical example for validation: (a) thermal analysis only, (b) mechanical analysis only, (c) thermo-mechanical analysis without and (d) with thermal expansion. Cases (a-c) can be achieved by setting $u_z$, $Q_r$ and $Q_m$, and $\alpha$ to zero, respectively. The results show good agreement between the proposed method and ABAQUS for all four cases, demonstrating the validity of the presented method in conducting individual and full analysis of finite-strain thermo-elastodynamics of soft tissues. Compared to the temperature field in (a), the temperature distribution in (c) was affected by the deformations and was further affected by thermal expansion in (d) which had a broader distribution of temperatures and a lower maximum temperature at the heat source centre. Compared with the displacement fields in (b) and (c), the displacement field in (d) had considerable local expansions at the three heat source regions that compressed the cylindrical holes and yielded less volume deformations. A statistical summary is presented in Table. 2.

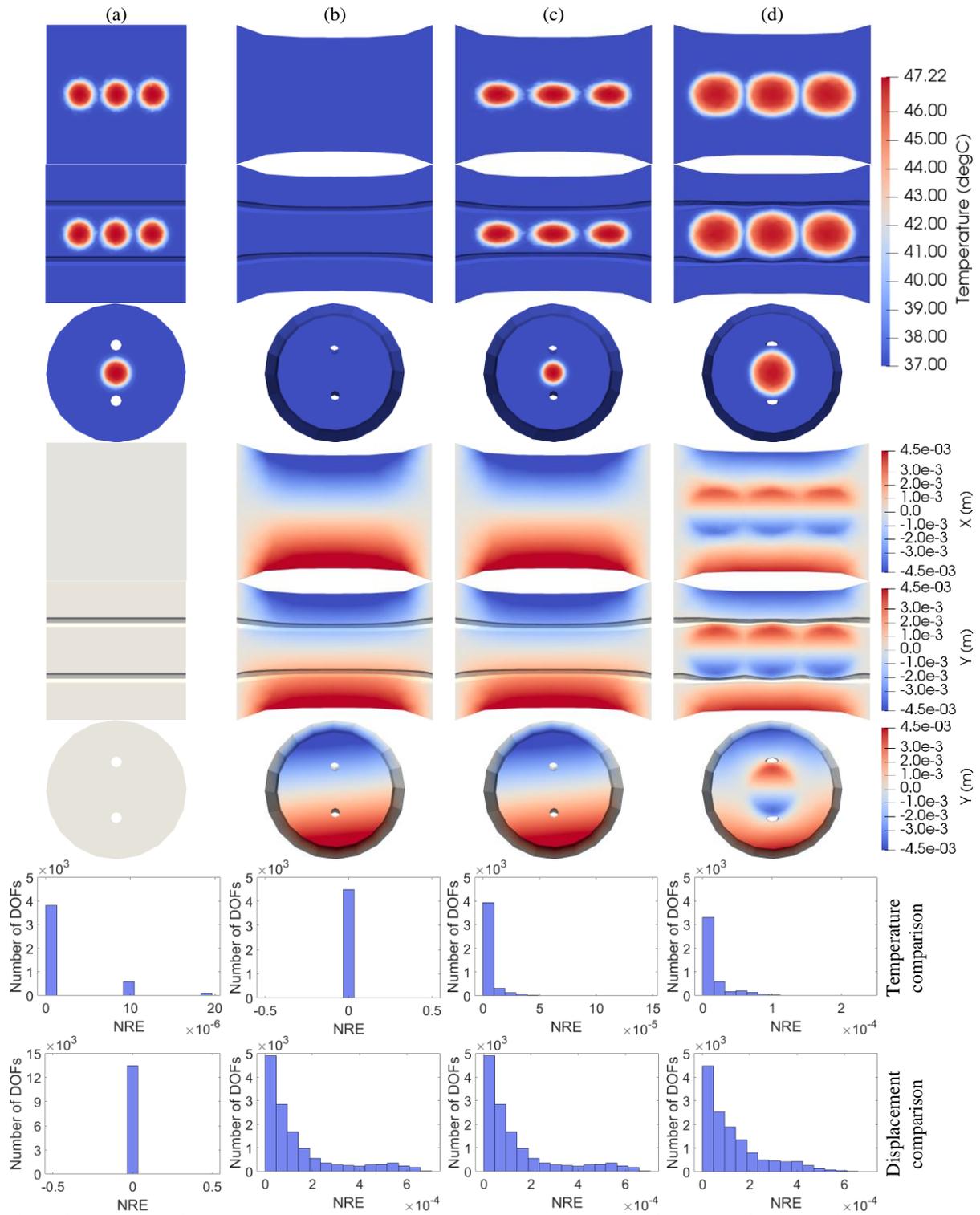

Fig. 4. Comparisons of temperatures, displacements, and numerical errors between four cases of the numerical example: (a) thermal analysis, (b) mechanical analysis, (c) thermo-mechanical analysis without and (d) with thermal expansion; rows #(1, 4), #(2, 5) and #(3, 6) shows the cross-section views of $xz$ plane, $yz$ plane, and $xy$ plane, respectively.

Table. 2 A statistical summary of the comparisons in Fig. 4.

| | | (a) Thermal | | (b) Mechanical | | (c) Thermo-mechanical without thermal expansion | | (d) Thermo-mechanical with thermal expansion | |
|---|---|---|---|---|---|---|---|---|---|
| | | P | A | P | A | P | A | P | A |
| Temperature comparison (°C) | Min | 37 | 37 | 37 | 37 | 37 | 37 | 37 | 37 |
| | Max | 47.2238 | 47.2236 | 37 | 37 | 47.2085 | 47.2084 | 47.1112 | 47.1111 |
| | Median | 37 | 37 | 37 | 37 | 37 | 37 | 37 | 37 |
| | Q1 | 37 | 37 | 37 | 37 | 37 | 37 | 37 | 37 |
| | Q3 | 37.4252 | 37.4252 | 37 | 37 | 37.4001 | 37.4002 | 37.7441 | 37.7437 |
| | $e_T$ | 1.2155e-6 | | 0 | | 3.1984e-6 | | 7.0775e-6 | |
| Displacement comparison (m) | Min | 0 | 0 | -0.0045 | -0.0045 | -0.0045 | -0.0045 | -0.0035 | -0.0035 |
| | Max | 0 | 0 | 0.0200 | 0.0200 | 0.0200 | 0.0200 | 0.0212 | 0.0212 |
| | Median | 0 | 0 | 3.6013e-4 | 3.6046e-4 | 3.6013e-4 | 3.6046e-4 | 3.2390e-4 | 3.2365e-4 |
| | Q1 | 0 | 0 | -1.2742e-4 | -1.2665e-4 | -1.2742e-4 | -1.2665e-4 | -1.4113e-4 | -1.4104e-4 |
| | Q3 | 0 | 0 | 0.0037 | 0.0037 | 0.0037 | 0.0037 | 0.0038 | 0.0038 |
| | $e_u$ | 0 | | 7.3887e-4 | | 7.3887e-4 | | 6.3711e-4 | |

P: Proposed, A: ABAQUS

Fig. 5 presents the effect of thermal expansion on the model temperature and displacement fields. As shown in Fig. 5(a-b), higher temperature regions were observed around the outer expanded region of the three heat sources, and even higher temperatures were observed between the heat sources due to the concentration of heat by the expansion of heat source regions. On the other hand, lower temperatures were observed around the middle section of the expanded regions, where there was less heat compared to the case without thermal expansion. As shown in Fig. 5(c-d), the effect of thermal expansion induced noticeable deformations around the heat source regions, and even greater deformations were observed at the cylindrical hole regions. On the other hand, fewer deformations were observed between heat sources as the regions expanded in opposite directions, resulting in less and close to zero resultant displacements.

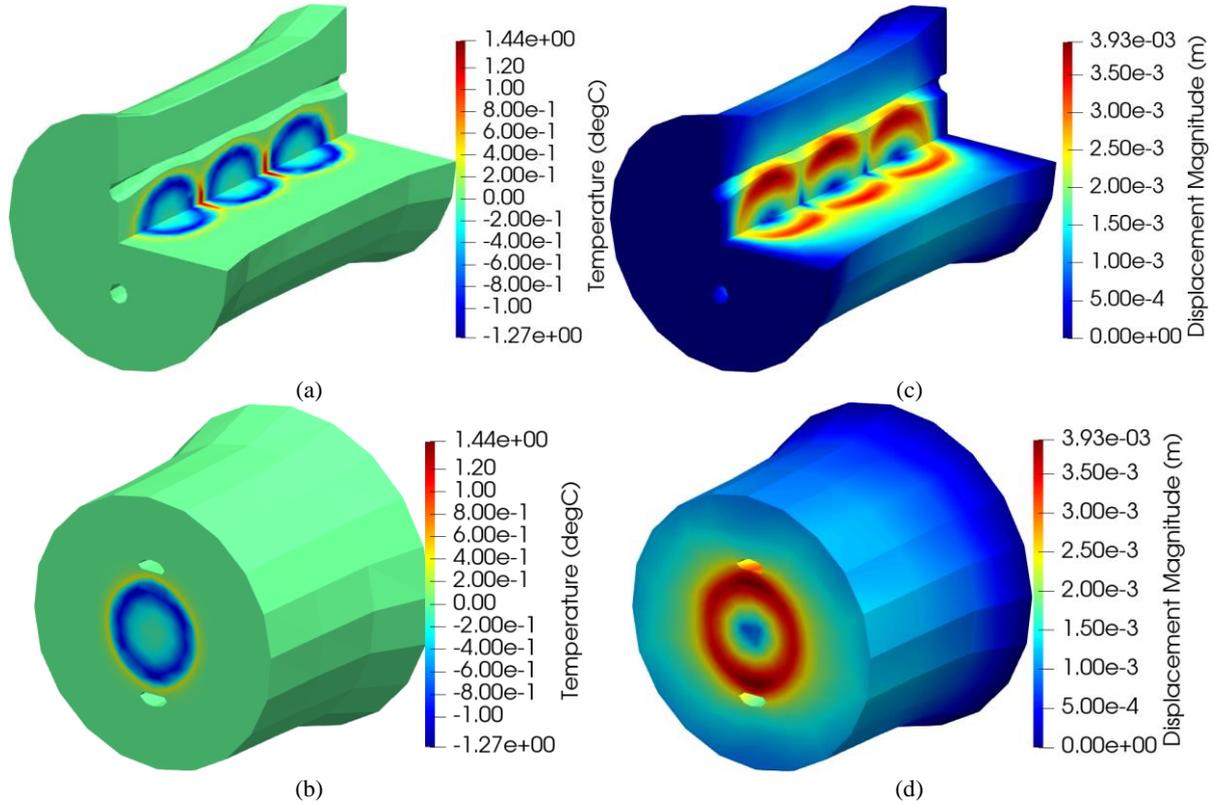

Fig. 5. The effect of thermal expansion on the temperature and displacement fields; colour maps indicate the difference in temperature and displacement values with and without thermal expansion.

## 6.2 CPU and GPU computational performance

The presented method was evaluated on an Intel(R) Core(TM) i5-2500K CPU @ 3.30 GHz with 8.0 GB RAM PC using Windows 10 operating system. Fig. 6(a) presents CPU solution times for a single time step in four-node linear tetrahedral (T4) and eight-node reduced-integrated hexahedral (H8) meshes for temperature-independent thermo-mechanical (TherMechTI), thermo-mechanical with thermal expansion (TherMechExpanTI), and temperature-dependent thermo-mechanical with thermal expansion (TherMechExpanTD) at 11 different model sizes. It can be seen that the computation times of TherMechTI were the least for both T4 and H8, followed by those of TherMechExpanTI and TherMechExpanTD. Furthermore, it was noticed that the solution times increased almost linearly with the increase of model sizes, and linear interpolation and extrapolation could be used to determine the solution times at unknown model sizes. Fig. 6(b) presents the computational overheads in CPU computation with the averaged overheads given in Table. 3. The computation time of the hourglass control algorithm [60] for H8 was considered in all evaluations.

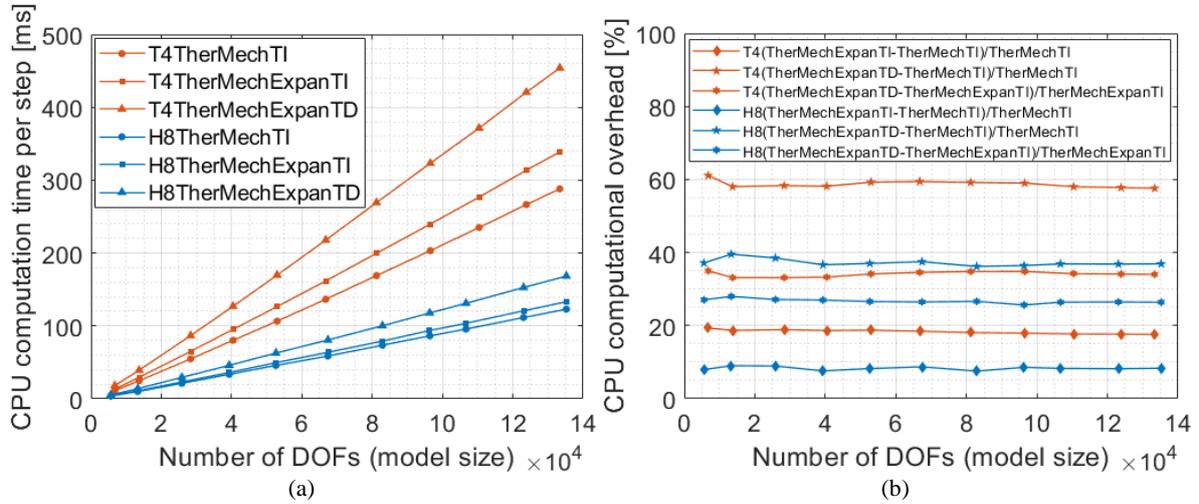

Fig. 6. (a) CPU solution times per simulation time step and (b) computational overheads in T4 and H8 meshes.

Table. 3 Averaged computational overheads in Fig. 6(b).

|  |  | T4 | H8 |
|---|---|---|---|
| Compared to TherMechTI | TherMechExpanTI | 18.35% | 8.31% |
|  | TherMechExpanTD | 58.72% | 37.25% |
| Compared to TherMechExpanTI | TherMechExpanTD | 34.12% | 26.71% |

The GPU implementation was executed on the same PC using an NVIDIA GeForce GTX 780 GPU with 2304 CUDA cores @ 863 MHz. The GPU computation was performed for the same model sizes and conditions as the CPU computation. Fig. 7(a-b) presents GPU solution times for a single time step in T4 and H8 for TherMechTI, TherMechExpanTI and TherMechExpanTD, and speed improvements (CPU/GPU time ratios). The GPU solution times exhibited a similar trend as the CPU times where they increased almost linearly with the increase of model sizes. Fig. 7(c-d) presents the computational overheads in GPU computation. A summary is presented in Table. 4.

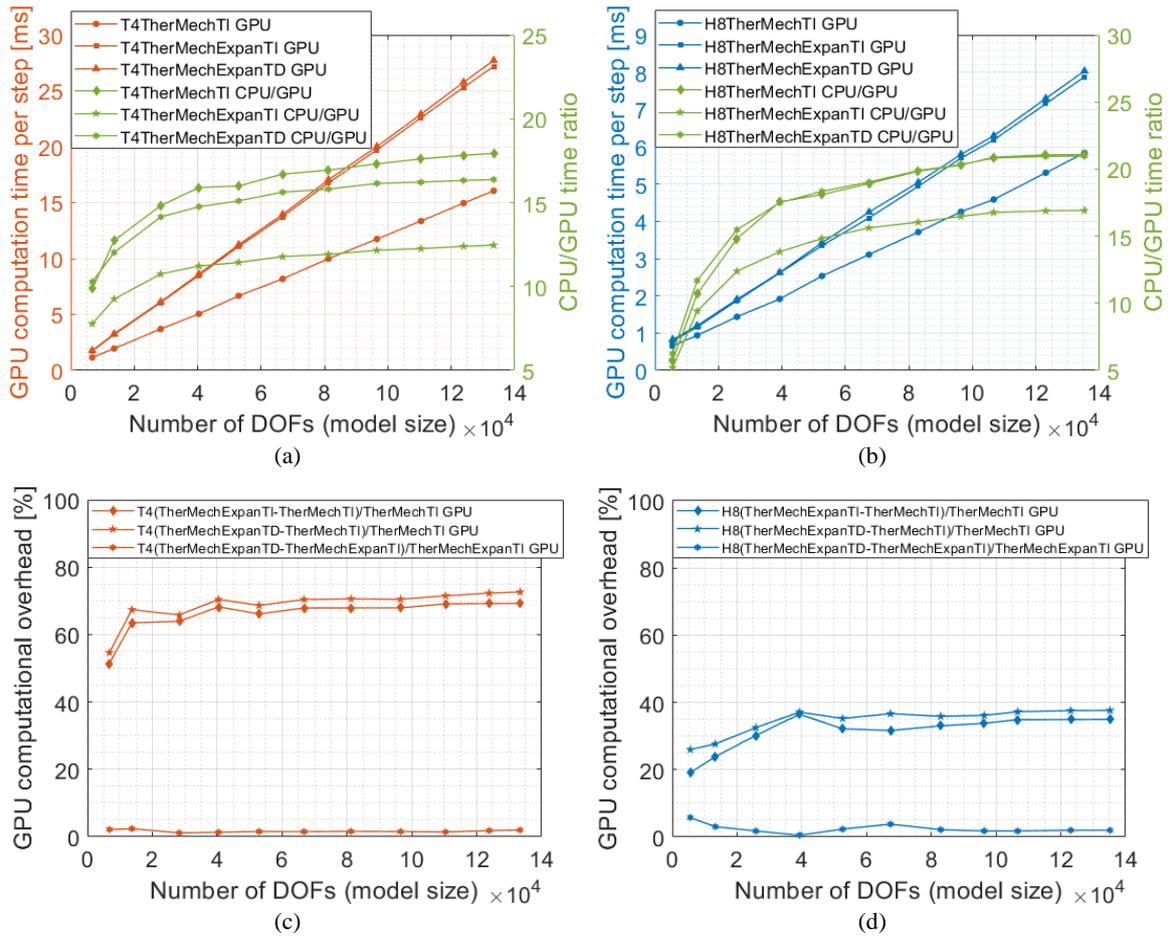

Fig. 7. GPU solution times per simulation time step and speed improvements (CPU/GPU) for (a) T4 and (b) H8 meshes, and (c-d) computational overheads.

Table. 4 GPU speed improvements over CPU and averaged computational overheads in Fig. 7.

|  |  | T4 | H8 |
|---|---|---|---|
| Maximum speed improvements | | | |
| Compared to CPU | TherMechTI | 17.94x | 21.08x |
|  | TherMechExpanTI | 12.46x | 16.92x |
|  | TherMechExpanTD | 16.38x | 20.97x |
| Computational overheads | | | |
| Compared to TherMechTI | TherMechExpanTI | 65.86% | 31.37% |
|  | TherMechExpanTD | 68.62% | 34.52% |
| Compared to TherMechExpanTI | TherMechExpanTD | 1.67% | 2.44% |

Fig. 8 presents the ratios of computation time of T4 and H8 in CPU and GPU executions for comparing the computational performance of finite elements. Linear interpolation was used to determine H8 computation times at T4 model sizes due to different mesh densities. At the same model sizes, T4 consumed between 2.3 and 2.8 times more computation time than those of H8 in CPU computation, and between 1.6 and 3.5 times in GPU computation.

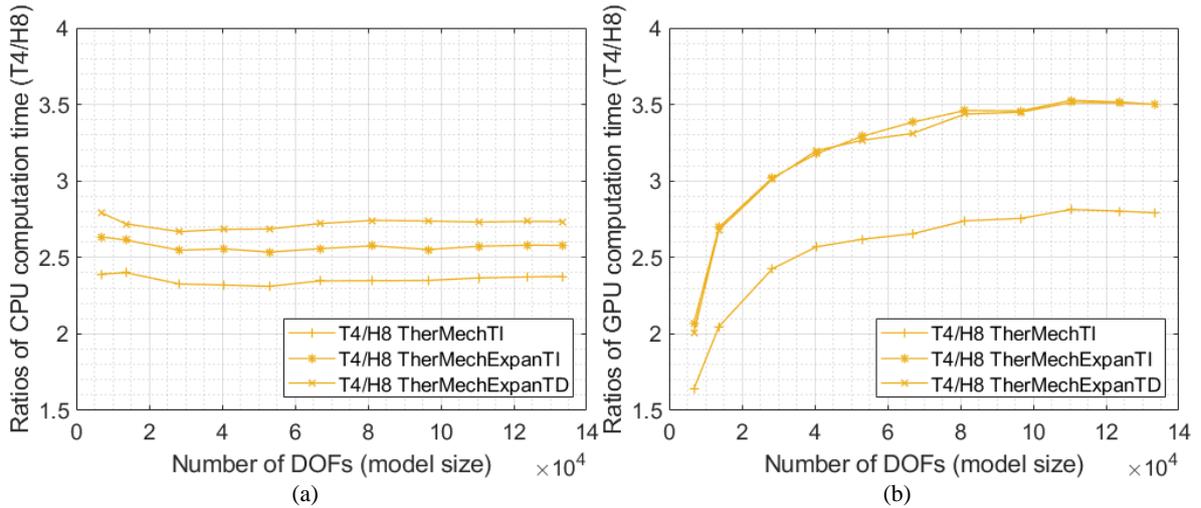

Fig. 8. Ratios of computation time of T4 over H8 in the (a) CPU and (b) GPU computation.

### 6.3 Bioheat-based temperature prediction and biomechanics-based image registration for surgical simulation of thermal ablation

The presented methodology is applied to simulate and predict soft tissue temperatures and deformations for patient-specific treatment planning of thermal ablation. Fig. 9 illustrates the workflow employed in the present study. Based on the acquired patient-specific medical image dataset, organ geometries were extracted by image segmentation to create 3D surface models for finite element meshing. Soft tissue thermal and mechanical material properties and load and boundary conditions were assigned to generate computational bioheat and biomechanics models for the temperature and displacement fields of soft tissues. The temperature field was used to predict the ablation zones (60 ℃ isotherms were used due to their strong relationship with the visible boundaries of coagulated tissues based on experimental observations [65]). The displacement field was used to incorporate soft tissue deformations in the medical images by image registration. The 60 ℃ isotherms and tissue deformations were overlaid and registered to the patient-specific image dataset to assist the operator of ablation for treatment planning and predicting treatment outcomes. Table. 5 presents the simulation settings. The medical image dataset was obtained from 3Dircadb 1.20 [66], and the 3D organ surface meshes were generated using 3D slicer [67] (http://www.slicer.org) and refined using MeshLab (http://meshlab.sourceforge.net). The finite element meshing was done using TetGen [68]. We refer the reader to Ref. [69] for a review of methods for generation of computational biomechanics models. The liver was modelled by a transversely isotropic neo-Hookean visco-hyperelastic model [64].

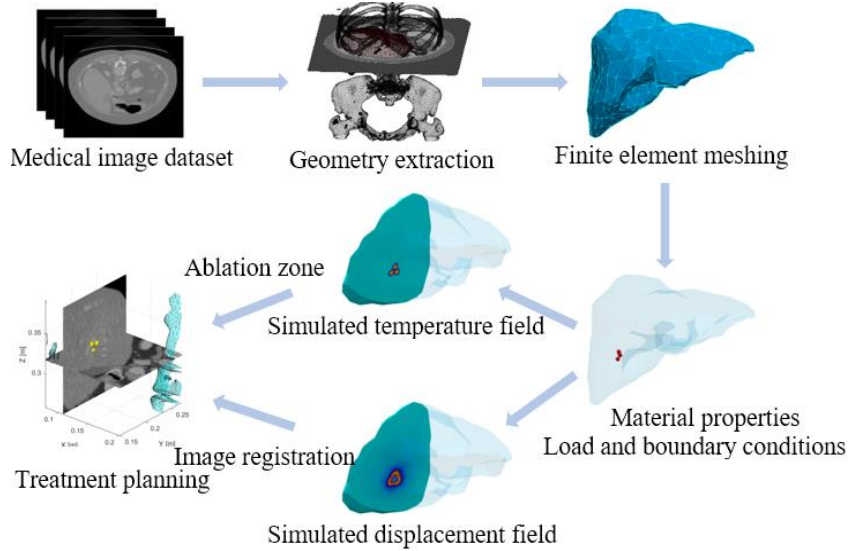

Fig. 9. The workflow used in the present work for treatment planning of thermal ablation.

Table. 5 Simulation settings of the liver model for temperature-dependent anisotropic thermo-visco-hyperelastic analysis.

| Geometry | Values | Unit | Description |
|---|---|---|---|
| Nodes | 3268 | | 13072 DOFs ($x, y, z, T$ per node) |
| Elements | 18007 | | Four-node linear tetrahedrons |
| Heat sources ×3 | $D = 0.01$ | [m] | Heat sources touch each other |
| Material properties | | | |
| Density $\rho$ | 1060 | [$kg/m^3$] | |
| Specific heat capacity $c$ | 3600 @ 37 ℃ | [$J/(kg \cdot ℃)$] | Temperature-dependent [45] |
| | 4300 @ 90 ℃ | [$J/(kg \cdot ℃)$] | Linear interpolation |
| Thermal conductivity $k$ | 0.53 @ 37 ℃ | [$W/(m \cdot ℃)$] | Temperature-dependent [45] |
| | 0.75 @ 90 ℃ | [$W/(m \cdot ℃)$] | Linear interpolation |
| Thermal expansion $\alpha$ | 0.0001 | [1/℃] | [70] |
| Transversely isotropic neo-Hookean hyperelastic model | | | $\Psi = \frac{\mu}{2}(\bar{I}_1 - 3) + \frac{\eta_a}{2}(\bar{I}_4 - 1)^2 + \frac{\kappa}{2}(J - 1)^2$ |
| Shear modulus $\mu$ | 1190.476 | [$Pa$] | Corresponding to Young's modulus 3500 Pa and Poisson's ratio |
| Bulk modulus $\kappa$ | 19444.444 | [$Pa$] | 0.47 [53] |
| Anisotropic coefficient $\eta_a$ | $\eta_a = 2\mu$ | [$Pa$] | [64] |
| Unit vector $^0\mathbf{a}$ | [1 0 0] | | To indicate local fibre directions for anisotropic behaviours |
| Prony terms $\varphi_1$ | 0.5 | | Viscoelastic behaviours [64] |
| Prony terms $\tau_1$ | 0.58 | | Viscoelastic behaviours [64] |
| Load and boundary conditions | | | |
| Heat source $Q_r$ | 9705360 | [$W/m^3$] | $Q_r = \rho SAR$ where $SAR = 1.5 \times 6.104 \times 10^3\ W/kg$ [70] |
| Metabolic heat generation $Q_m$ | 33800 | [$W/m^3$] | [70] |
| Blood perfusion rate $w_b$ | 26.6 | [$kg/(m^3 \cdot s)$] | [21] |
| Blood specific heat capacity $c_b$ | 3617 | [$J/(kg \cdot K)$] | [21] |
| Arterial blood temperature $T_a$ | 37 | [℃] | |
| Exterior of the domain: adiabatic boundary condition | | | |
| Simulation time | 25 s | Time step | 0.0002 s (125000 steps) |

Fig. 10 presents the simulated temperatures, 60 ℃ isotherms, and displacement fields in the liver using TherMechTI, TherMechExpanTI and TherMechExpanTD simulations, and Table. 6 presents the results of maximum temperatures and maximum and minimum displacements. The temperature-independent simulation employed the specific heat and conductivity values at 37 ℃. The TherMechExpanTD had the lowest maximum temperature, followed by those of TherMechExpanTI and TherMechTI. Due to higher tissue temperatures, TherMechExpanTI had a more pronounced thermal expansion than TherMechExpanTD. In all, TherMechExpanTD had the smallest predicted ablation volumes. Fig. 11 shows the combined results where the predicted ablation zones are displayed in the registered patient-specific Computed Tomography (CT) images considering soft tissue deformations for the operator of ablation to conduct treatment planning and predict treatment outcomes.

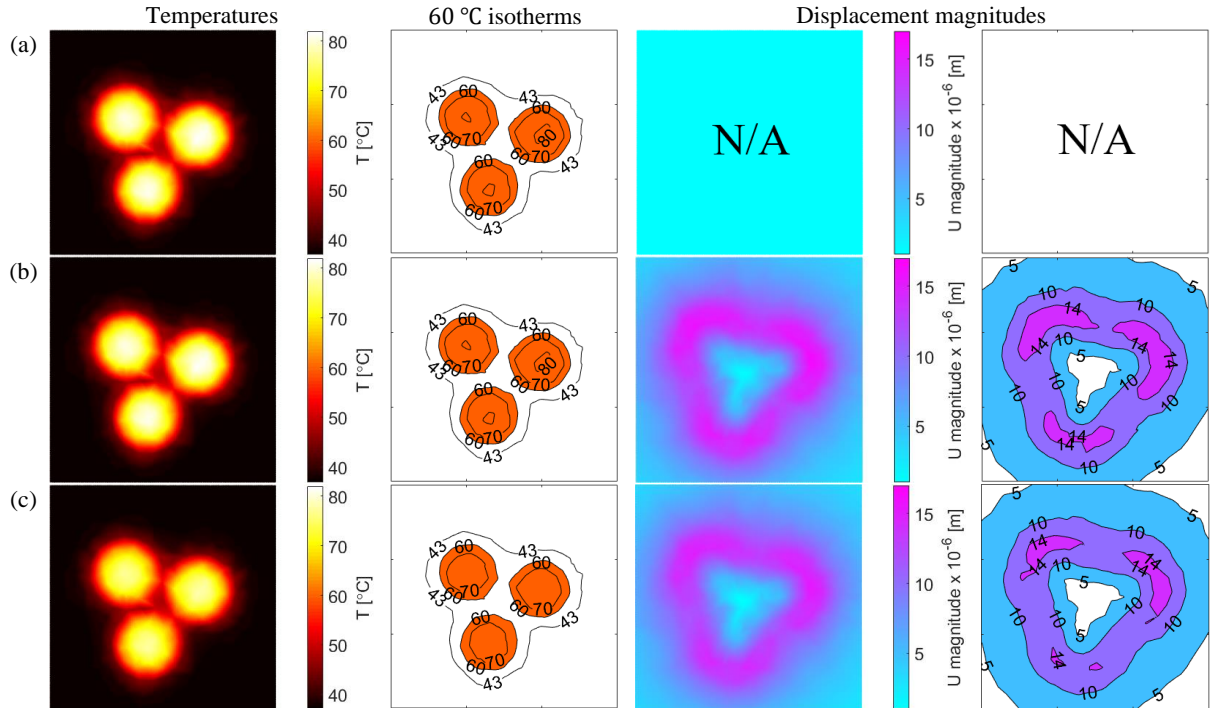

Fig. 10. Simulation results of inducing three heat sources in the liver using (a) TherMechTI, (b) TherMechExpanTI and (c) TherMechExpanTD simulations (note: (a) TherMechTI does not involve thermal expansion; hence, N/A is used in "Displacement magnitudes").

Table. 6 Temperatures and displacements in the liver in Fig. 10 to show the effects of thermal expansion and temperature-in/dependent material properties.

| | $T_{max}$ (°C) | $u_{x,max}$ (m) | $u_{y,max}$ (m) | $u_{z,max}$ (m) | $u_{x,min}$ (m) | $u_{y,min}$ (m) | $u_{z,min}$ (m) |
|---|---|---|---|---|---|---|---|
| TherMechTI | 81.8196 | 0 | 0 | 0 | 0 | 0 | 0 |
| TherMechExpanTI | 81.7794 | 1.25e-5 | 1.65e-5 | 1.55e-5 | -1.19e-5 | -1.50e-5 | -1.44e-5 |
| TherMechExpanTD | 78.2727 | 1.17e-5 | 1.55e-5 | 1.46e-5 | -1.11e-5 | -1.42e-5 | -1.36e-5 |

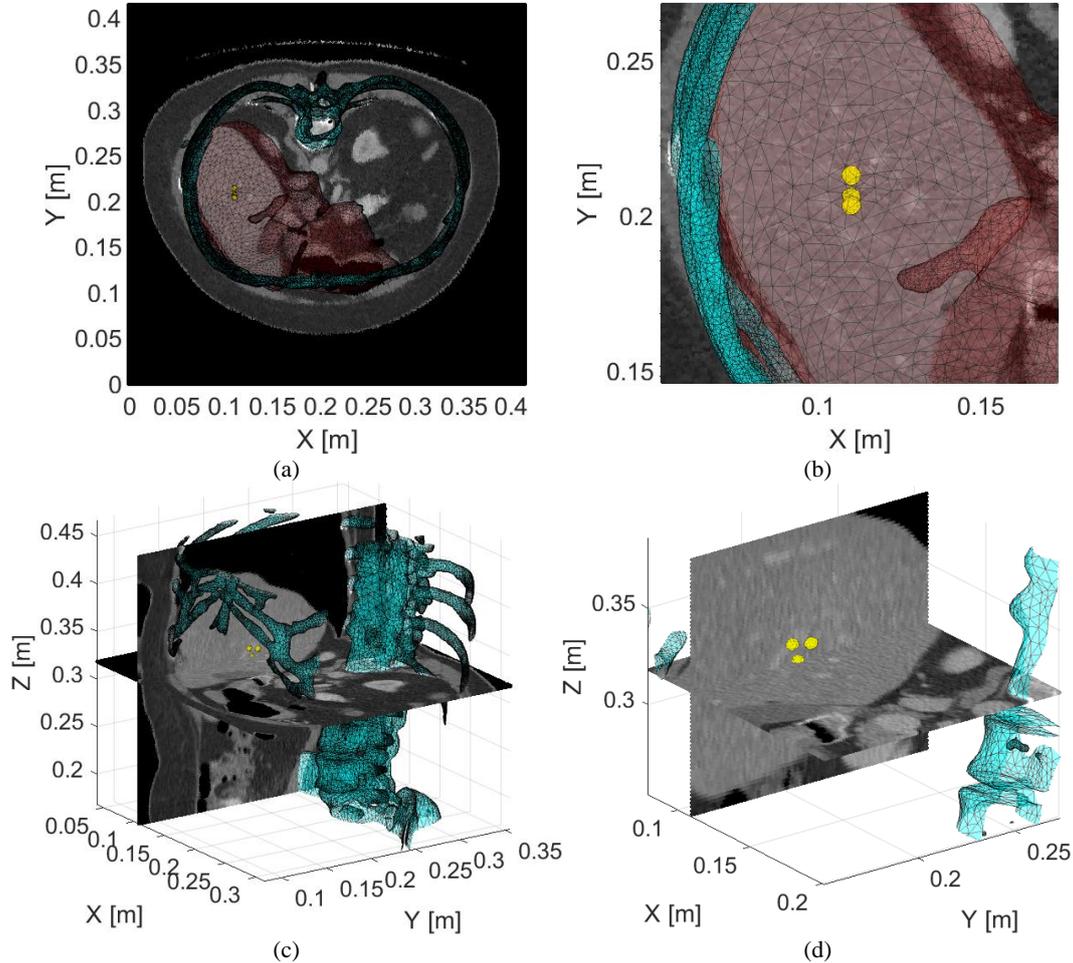

Fig. 11. The simulated 60 ℃ isotherms using TherMechExpanTD are visualised on the image registered patient-specific CT images from different views.

## 7. Discussions

*Importance of biomechanics modelling*: Biomechanics modelling provides the basis for understanding of many thermo-mechanical behaviours during thermal ablation. In percutaneous RFA, the liver experiences deformations due to needle insertion. In MWA, soft tissues experience expansions close to the antenna's active tip [11], and an overall tissue shrinkage at the end of the procedure [12]. In other non-invasive thermal ablation treatments such as HIFU ablation, target organs such as the liver and kidneys experience respiration-induced movements. These soft tissue behaviours require the coupled computational bioheat transfer and biomechanics modelling, instead of bioheat transfer alone which was used in many existing works. This will improve the understanding of parametric dependence of thermal doses and exposure outcomes while the target tissues are in dynamics.

*Applicability to other hyperthermia treatments*: Although the clinically relevant application is demonstrated by a simulation of thermal ablation for local hyperthermia, the principle of the coupled bioheat and biomechanics modelling is generic and can be straightforwardly adapted for regional and whole-body hyperthermia simulations, given the regional and whole-body finite element meshes, material properties, and load and boundary conditions.

*Heat source based on modality*: In the present work, we used spherical heat source regions with a uniform heat value (similar to Yuan [71]) to simplify the problem for the purpose of algorithm verification. However, the actual heat deposition is usually not uniform and should be determined based on the specific type of modality used for thermal ablation. For instance, it needs to solve the propagation of ultrasound in soft tissues for acoustic pressures for volumetric heat depositions in HIFU [72], Maxwell's equation for the electromagnetic field in MWA [11] and RFA [73], and laser intensity described by Beer-Lambert's law in laser ablation [23]. A review of heat source generated by different applications can be found in [74].

*Thermal damage of tissues*: The extent of thermal damage in tissues can be estimated based on (i) temperature thresholding, (ii) the cumulative equivalent minutes (CEM) at 43 °C, or (iii) an Arrhenius damage integral [46]. In the present work, we used the temperature thresholding based on 60 °C isotherms owing to their strong relationship with the visible boundaries of coagulated tissues based on experimental observations [65]. This is also supported by numerical estimations from the CEM43 and the Arrhenius integral, showing that the tissues will be necrotised almost instantly when the temperature is 60 °C.

*Anisotropy*: The presented method accommodates anisotropic and temperature-dependent properties of soft tissues. Anisotropic thermal conductivity, anisotropic thermal expansion coefficient, and anisotropic hyperelasticity can be incorporated into the constitutive models for soft tissue temperature and deformation computation. Temperature-dependent specific heat capacity and thermal conductivity can also be done similarly.

*Computational performance*: The computational performance is mainly affected by two factors, the total number of time steps and the number of DOFs. These two factors are controlled by the total simulation time, time step size, and model discretization. A longer simulation time usually yields longer computation time. This can be reduced by using a larger time step size to decrease the total number of time steps; however, the maximum allowable time step size is limited by a critical value in the explicit time integration [75] which is related to the spatial discretization and material properties. With given material properties, a more refined discretization usually leads to a smaller critical time step, which is limited by the smallest element in the mesh, and also leads to an increase in number of DOFs, further increasing computational cost. Therefore, simulations need to consider the mesh discretization for desired efficiency. Some methods such as deep learning [76] and reduce order modelling [77] were reported to permit a larger time step size to be used for more efficient computation.

*Finite element considerations*: finally, considerations must also be given to the type of finite elements (T4 or H8) used for simulation. As shown in Fig. 8, T4 meshes are between 1.6 and 3.5 times computationally more expensive than H8 with the same number of DOFs; however, the generation of H8 meshes is typically more difficult due to the complicated irregular shape of body organs, requiring significant time and manual interventions to complete a single mesh [69]. In contrast, the key advantage of T4 meshes is that they can be generated automatically based on the surface geometry of patient-specific anatomy, and this is often a standard feature of many mesh processing packages, such as TetGen [68] mentioned previously.

## 8. Conclusion

Simulation of thermal ablation requires the coupled computational bioheat transfer and biomechanics modelling for accurate prediction of soft tissue temperatures. We achieve this by presenting a thermo-visco-hyperelastic total Lagrangian explicit dynamics finite element algorithm. The essential advantage of the presented method is that it enables full nonlinear modelling of the anisotropic, finite-strain, temperature-dependent, thermal, and viscoelastic behaviours of soft tissues, instead of the linear elastic, linear viscoelastic, and thermal-only modelling in the existing works. To achieve simulations in real-time, the presented algorithm is developed to be well suited for GPU parallel computation. We demonstrate a clinically relevant scenario using a simulation of thermal ablation in the liver. Future works will be devoted to the application of the presented method in simulation of the planning, guidance and training of thermal ablation and other hyperthermia treatments and the development of computer-assisted treatment systems for personalised thermal ablation.

**Acknowledgment**

This work is funded by the National Health and Medical Research Council (NHMRC), Australia, Grant [1093314].

**References**

[1] A. Andreozzi, L. Brunese, M. Iasiello, C. Tucci, and G. P. Vanoli, "Modeling Heat Transfer in Tumors: A Review of Thermal Therapies," *Ann Biomed Eng,* vol. 47, no. 3, pp. 676-693, 2019.
[2] D. J. Breen, and R. Lencioni, "Image-guided ablation of primary liver and renal tumours," *Nat Rev Clin Oncol,* vol. 12, no. 3, pp. 175-86, 2015.
[3] D. S. Lu, S. S. Raman, P. Limanond, D. Aziz, J. Economou, R. Busuttil, and J. Sayre, "Influence of large peritumoral vessels on outcome of radiofrequency ablation of liver tumors," *J Vasc Interv Radiol,* vol. 14, no. 10, pp. 1267-74, 2003.
[4] C. Boutros, P. Somasundar, S. Garrean, A. Saied, and N. J. Espat, "Microwave coagulation therapy for hepatic tumors: review of the literature and critical analysis," *Surg Oncol,* vol. 19, no. 1, pp. e22-32, 2010.
[5] J. Yu, P. Liang, X. Yu, F. Liu, L. Chen, and Y. Wang, "A comparison of microwave ablation and bipolar radiofrequency ablation both with an internally cooled probe: results in ex vivo and in vivo porcine livers," *European journal of radiology,* vol. 79, no. 1, pp. 124-130, 2011.
[6] O. Seror, "Ablative therapies: advantages and disadvantages of radiofrequency, cryotherapy, microwave and electroporation methods, or how to choose the right method for an individual patient?," *Diagnostic and interventional imaging,* vol. 96, no. 6, pp. 617-624, 2015.
[7] N. V. Violi, R. Duran, B. Guiu, J.-P. Cercueil, C. Aubé, A. Digklia, I. Pache, P. Deltenre, J.-F. Knebel, and A. Denys, "Efficacy of microwave ablation versus radiofrequency ablation for the treatment of hepatocellular carcinoma in patients with chronic liver disease: a randomised controlled phase 2 trial," *The Lancet Gastroenterology & Hepatology,* vol. 3, no. 5, pp. 317-325, 2018.
[8] L. C. Bertot, M. Sato, R. Tateishi, H. Yoshida, and K. Koike, "Mortality and complication rates of percutaneous ablative techniques for the treatment of liver tumors: a systematic review," *European radiology,* vol. 21, no. 12, pp. 2584-2596, 2011.

[9] Y. Fang, W. Chen, X. Liang, D. Li, H. Lou, R. Chen, K. Wang, and H. Pan, "Comparison of long-term effectiveness and complications of radiofrequency ablation with hepatectomy for small hepatocellular carcinoma," *J Gastroenterol Hepatol,* vol. 29, no. 1, pp. 193-200, 2014.

[10] M. Moche, H. Busse, J. J. Futterer, C. A. Hinestrosa, D. Seider, P. Brandmaier, M. Kolesnik, S. Jenniskens, R. Blanco Sequeiros, G. Komar, M. Pollari, M. Eibisberger, H. R. Portugaller, P. Voglreiter, R. Flanagan, P. Mariappan, and M. Reinhardt, "Clinical evaluation of in silico planning and real-time simulation of hepatic radiofrequency ablation (ClinicIMPPACT Trial)," *Eur Radiol,* vol. 30, no. 2, pp. 934-942, 2020.

[11] V. Lopresto, R. Pinto, L. Farina, and M. Cavagnaro, "Treatment planning in microwave thermal ablation: clinical gaps and recent research advances," *Int J Hyperthermia,* vol. 33, no. 1, pp. 83-100, 2017.

[12] L. Farina, Y. Nissenbaum, M. Cavagnaro, and S. N. Goldberg, "Tissue shrinkage in microwave thermal ablation: comparison of three commercial devices," *International Journal of Hyperthermia,* vol. 34, no. 4, pp. 382-391, 2018.

[13] A. Andreozzi, M. Iasiello, and P. A. Netti, "A thermoporoelastic model for fluid transport in tumour tissues," *J R Soc Interface,* vol. 16, no. 154, pp. 20190030, 2019.

[14] S. Chung, and K. Vafai, "Mechanobiology of low-density lipoprotein transport within an arterial wall—impact of hyperthermia and coupling effects," *Journal of biomechanics,* vol. 47, no. 1, pp. 137-147, 2014.

[15] M. Iasiello, K. Vafai, A. Andreozzi, and N. Bianco, "Low-density lipoprotein transport through an arterial wall under hyperthermia and hypertension conditions–An analytical solution," *Journal of biomechanics,* vol. 49, no. 2, pp. 193-204, 2016.

[16] V. Suomi, J. Jaros, B. Treeby, and R. O. Cleveland, "Full Modeling of High-Intensity Focused Ultrasound and Thermal Heating in the Kidney Using Realistic Patient Models," *IEEE Trans Biomed Eng,* vol. 65, no. 11, pp. 2660-2670, 2018.

[17] K. Sumser, E. Neufeld, R. F. Verhaart, V. Fortunati, G. M. Verduijn, T. Drizdal, T. van Walsum, J. F. Veenland, and M. M. Paulides, "Feasibility and relevance of discrete vasculature modeling in routine hyperthermia treatment planning," *Int J Hyperthermia,* vol. 36, no. 1, pp. 801-811, 2019.

[18] J. Ma, X. Yang, Y. Sun, and J. Yang, "Thermal damage in three-dimensional vivo bio-tissues induced by moving heat sources in laser therapy," *Sci Rep,* vol. 9, no. 1, pp. 10987, 2019.

[19] K. Miller, "Computational Biomechanics for Patient-Specific Applications," *Ann Biomed Eng,* vol. 44, no. 1, pp. 1-2, 2016.

[20] Y. Tong, "High precision solution for thermo-elastic equations using stable node-based smoothed finite element method," *Applied Mathematics and Computation,* vol. 336, pp. 272-287, 2018.

[21] W. Karaki, Rahul, C. A. Lopez, D. A. Borca-Tasciuc, and S. De, "A continuum thermomechanical model of in vivo electrosurgical heating of hydrated soft biological tissues," *Int J Heat Mass Transf,* vol. 127, no. Pt A, pp. 961-974, 2018.

[22] T. Kröger, I. Altrogge, T. Preusser, P. L. Pereira, D. Schmidt, A. Weihusen, and H.-O. Peitgen, "Numerical Simulation of Radio Frequency Ablation with State Dependent Material Parameters in Three Space Dimensions," *Medical Image Computing and Computer-Assisted Intervention – MICCAI 2006.* pp. 380-388.

[23] P. Wongchadakul, P. Rattanadecho, and T. Wessapan, "Implementation of a thermomechanical model to simulate laser heating in shrinkage tissue (effects of wavelength, laser irradiation intensity, and irradiation beam area)," *International Journal of Thermal Sciences,* vol. 134, pp. 321-336, 2018.

[24] X. Li, Q.-H. Qin, and X. Tian, "Thermo-viscoelastic analysis of biological tissue during hyperthermia treatment," *Applied Mathematical Modelling,* vol. 79, pp. 881-895, 2020.

[25] H. H. Pennes, "Analysis of tissue and arterial blood temperatures in the resting human forearm," *J Appl Physiol,* vol. 1, no. 2, pp. 93-122, 1948.

[26] W. Wulff, "The energy conservation equation for living tissue," *IEEE Trans Biomed Eng*, no. 6, pp. 494-495, 1974.

[27] H. G. Klinger, "Heat transfer in perfused biological tissue. I. General theory," *Bull Math Biol,* vol. 36, no. 4, pp. 403-15, 1974.

[28] M. M. Chen, and K. R. Holmes, "Microvascular contributions in tissue heat transfer," *Ann N Y Acad Sci,* vol. 335, no. 1, pp. 137-50, 1980.

[29] A. Nakayama, and F. Kuwahara, "A general bioheat transfer model based on the theory of porous media," *International Journal of Heat and Mass Transfer,* vol. 51, no. 11-12, pp. 3190-3199, 2008.

[30] A. Kotte, G. van Leeuwen, J. de Bree, J. van der Koijk, H. Crezee, and J. Lagendijk, "A description of discrete vessel segments in thermal modelling of tissues," *Phys Med Biol,* vol. 41, no. 5, pp. 865-84, 1996.

[31] D. Y. Tzou, *Macro-to microscale heat transfer: the lagging behavior*: John Wiley & Sons, 2014.

[32] A. Bhowmik, R. Singh, R. Repaka, and S. C. Mishra, "Conventional and newly developed bioheat transport models in vascularized tissues: A review," *Journal of Thermal Biology,* vol. 38, no. 3, pp. 107-125, 2013.

[33] P. Gupta, and A. Srivastava, "Numerical analysis of thermal response of tissues subjected to high intensity focused ultrasound," *Int J Hyperthermia,* vol. 35, no. 1, pp. 419-434, 2018.

[34] J. Zhang, and S. Chauhan, "Neural network methodology for real-time modelling of bio-heat transfer during thermo-therapeutic applications," *Artif Intell Med,* vol. 101, pp. 101728, 2019.

[35] J. Zhang, J. Hills, Y. Zhong, B. Shirinzadeh, J. Smith, and C. Gu, "Modeling of soft tissue thermal damage based on GPU acceleration," *Comput Assist Surg (Abingdon),* vol. 24, no. sup1, pp. 5-12, 2019.

[36] M. C. Kolios, A. E. Worthington, M. D. Sherar, and J. W. Hunt, "Experimental evaluation of two simple thermal models using transient temperature analysis," *Phys Med Biol,* vol. 43, no. 11, pp. 3325, 1998.

[37] E. H. Wissler, "Pennes' 1948 paper revisited," *J Appl Physiol (1985),* vol. 85, no. 1, pp. 35-41, 1998.

[38] M. A. Solovchuk, M. Thiriet, and T. W. H. Sheu, "Computational study of acoustic streaming and heating during acoustic hemostasis," *Applied Thermal Engineering,* vol. 124, pp. 1112-1122, 2017.

[39] M. Ge, K. Chua, C. Shu, and W. Yang, "Analytical and numerical study of tissue cryofreezing via the immersed boundary method," *International Journal of Heat and Mass Transfer,* vol. 83, pp. 1-10, 2015.

[40] Y. Zhang, "Generalized dual-phase lag bioheat equations based on nonequilibrium heat transfer in living biological tissues," *International Journal of Heat and Mass Transfer,* vol. 52, no. 21-22, pp. 4829-4834, 2009.

[41] F. Xu, K. Seffen, and T. Lu, "Non-Fourier analysis of skin biothermomechanics," *International Journal of Heat and Mass Transfer,* vol. 51, no. 9-10, pp. 2237-2259, 2008.

[42] F. Xu, T. Lu, and K. Seffen, "Biothermomechanical behavior of skin tissue," *Acta Mechanica Sinica,* vol. 24, no. 1, pp. 1-23, 2008.

[43] G. C. Bourantas, M. Ghommem, G. C. Kagadis, K. Katsanos, V. C. Loukopoulos, V. N. Burganos, and G. C. Nikiforidis, "Real-time tumor ablation simulation based on the dynamic mode decomposition method," *Med Phys,* vol. 41, no. 5, pp. 053301, 2014.

[44] J. Zhang, J. Hills, Y. Zhong, B. Shirinzadeh, J. Smith, and C. Gu, "Temperature-Dependent Thermomechanical Modeling of Soft Tissue Deformation," *Journal of Mechanics in Medicine and Biology,* vol. 18, no. 08, pp. 1840021, 2019.

[45] S. R. Guntur, K. I. Lee, D. G. Paeng, A. J. Coleman, and M. J. Choi, "Temperature-dependent thermal properties of ex vivo liver undergoing thermal ablation," *Ultrasound Med Biol,* vol. 39, no. 10, pp. 1771-84, 2013.

[46] M. Schwenke, J. Georgii, and T. Preusser, "Fast Numerical Simulation of Focused Ultrasound Treatments During Respiratory Motion With Discontinuous Motion Boundaries," *IEEE Trans Biomed Eng,* vol. 64, no. 7, pp. 1455-1468, 2017.
[47] Y.-C. Fung, *Biomechanics: Mechanical Properties of Living Tissues*: Springer-Verlag, 1993.
[48] J. A. Weiss, B. N. Maker, and S. Govindjee, "Finite element implementation of incompressible, transversely isotropic hyperelasticity," *Computer Methods in Applied Mechanics and Engineering,* vol. 135, no. 1-2, pp. 107-128, 1996.
[49] J. Zhang, Y. Zhong, and C. Gu, "Deformable Models for Surgical Simulation: A Survey," *IEEE Rev Biomed Eng,* vol. 11, pp. 143-164, 2018.
[50] G. A. Holzapfel, "On Large Strain Viscoelasticity: Continuum Formulation and Finite Element Applications to Elastomeric Structures," *International Journal for Numerical Methods in Engineering,* vol. 39, no. 22, pp. 3903-3926, 1996.
[51] A. G. Holzapfel, *Nonlinear solid mechanics II*, 2000.
[52] L. Vujosevic, and V. Lubarda, "Finite-strain thermoelasticity based on multiplicative decomposition of deformation gradient," *THEORETICAL AND APPLIED MECHANICS,* vol. 28, no. 29, pp. 379-399, 2002.
[53] J. Zhang, and S. Chauhan, "Fast computation of soft tissue thermal response under deformation based on fast explicit dynamics finite element algorithm for surgical simulation," *Comput Methods Programs Biomed,* vol. 187, pp. 105244, 2020.
[54] V. A. Lubarda, "Constitutive theories based on the multiplicative decomposition of deformation gradient: Thermoelasticity, elastoplasticity, and biomechanics," *Applied Mechanics Reviews,* vol. 57, no. 2, pp. 95-108, 2004.
[55] J. Zhang, and S. Chauhan, "Fast explicit dynamics finite element algorithm for transient heat transfer," *International Journal of Thermal Sciences,* vol. 139, pp. 160-175, 2019.
[56] J. Zhang, and S. Chauhan, "Real-time computation of bio-heat transfer in the fast explicit dynamics finite element algorithm (FED-FEM) framework," *Numerical Heat Transfer, Part B: Fundamentals,* vol. 75, no. 4, pp. 217-238, 2019.
[57] K. Miller, G. Joldes, D. Lance, and A. Wittek, "Total Lagrangian explicit dynamics finite element algorithm for computing soft tissue deformation," *Communications in Numerical Methods in Engineering,* vol. 23, no. 2, pp. 121-134, 2006.
[58] K.-J. Bathe, *Finite element procedures*, 2006.
[59] G. Szekely, C. Brechbuhler, R. Hutter, A. Rhomberg, N. Ironmonger, and P. Schmid, "Modelling of soft tissue deformation for laparoscopic surgery simulation," *Med Image Anal,* vol. 4, no. 1, pp. 57-66, 2000.
[60] G. R. Joldes, A. Wittek, and K. Miller, "An efficient hourglass control implementation for the uniform strain hexahedron using the Total Lagrangian formulation," *Communications in Numerical Methods in Engineering,* vol. 24, no. 11, pp. 1315-1323, 2007.
[61] A. Nealen, M. Müller, R. Keiser, E. Boxerman, and M. Carlson, "Physically Based Deformable Models in Computer Graphics," *Computer Graphics Forum,* vol. 25, no. 4, pp. 809-836, 2006.
[62] S. Cotin, H. Delingette, and N. Ayache, "A hybrid elastic model for real-time cutting, deformations, and force feedback for surgery training and simulation," *The Visual Computer,* vol. 16, no. 8, pp. 437-452, 2000.
[63] Z. A. Taylor, M. Cheng, and S. Ourselin, "High-speed nonlinear finite element analysis for surgical simulation using graphics processing units," *IEEE Trans Med Imaging,* vol. 27, no. 5, pp. 650-63, 2008.
[64] Z. A. Taylor, O. Comas, M. Cheng, J. Passenger, D. J. Hawkes, D. Atkinson, and S. Ourselin, "On modelling of anisotropic viscoelasticity for soft tissue simulation: numerical solution and GPU execution," *Med Image Anal,* vol. 13, no. 2, pp. 234-44, 2009.
[65] P. Prakash, "Theoretical modeling for hepatic microwave ablation," *The Open Biomedical Engineering Journal,* vol. 4, pp. 27-38, 2010.
[66] https://www.ircad.fr/research/3d-ircadb-01/, "3D-IRCADb-01-20," IRCAD France.
[67] A. Fedorov, R. Beichel, J. Kalpathy-Cramer, J. Finet, J.-C. Fillion-Robin, S. Pujol, C. Bauer, D. Jennings, F. Fennessy, and M. Sonka, "3D Slicer as an image computing platform for the Quantitative Imaging Network," *Magnetic resonance imaging,* vol. 30, no. 9, pp. 1323-1341, 2012.
[68] H. Si, "TetGen, a Delaunay-Based Quality Tetrahedral Mesh Generator," *ACM Transactions on Mathematical Software,* vol. 41, no. 2, pp. 1-36, 2015.
[69] A. Wittek, N. M. Grosland, G. R. Joldes, V. Magnotta, and K. Miller, "From Finite Element Meshes to Clouds of Points: A Review of Methods for Generation of Computational Biomechanics Models for Patient-Specific Applications," *Ann Biomed Eng,* vol. 44, no. 1, pp. 3-15, 2016.
[70] P. Rattanadecho, and P. Keangin, "Numerical study of heat transfer and blood flow in two-layered porous liver tissue during microwave ablation process using single and double slot antenna," *International Journal of Heat and Mass Transfer,* vol. 58, no. 1-2, pp. 457-470, 2013.
[71] P. Yuan, "Numerical analysis of temperature and thermal dose response of biological tissues to thermal non-equilibrium during hyperthermia therapy," *Med Eng Phys,* vol. 30, no. 2, pp. 135-43, 2008.
[72] J. Zhang, N.-D. Bui, W. Cheung, S. K. Roberts, and S. Chauhan, "Fast computation of desired thermal dose: Application to focused ultrasound-induced lesion planning," *Numerical Heat Transfer, Part A: Applications,* vol. 77, no. 6, pp. 666-682, 2020.
[73] B. Prasad, J. K. Kim, and S. Kim, "Role of Simulations in the Treatment Planning of Radiofrequency Hyperthermia Therapy in Clinics," *J Oncol,* vol. 2019, pp. 9685476, 2019.
[74] A. Andreozzi, M. Iasiello, and C. Tucci, "An overview of mathematical models and modulated-heating protocols for thermal ablation," 2020.
[75] R. Courant, K. Friedrichs, and H. Lewy, "On the partial difference equations of mathematical physics," *IBM journal of Research and Development,* vol. 11, no. 2, pp. 215-234, 1967.
[76] F. Meister, T. Passerini, V. Mihalef, A. Tuysuzoglu, A. Maier, and T. Mansi, "Deep learning acceleration of Total Lagrangian Explicit Dynamics for soft tissue mechanics," *Computer Methods in Applied Mechanics and Engineering,* vol. 358, pp. 112628, 2020.
[77] Z. A. Taylor, S. Crozier, and S. Ourselin, "A reduced order explicit dynamic finite element algorithm for surgical simulation," *IEEE Trans Med Imaging,* vol. 30, no. 9, pp. 1713-21, 2011.